\documentclass[traditabstract]{aa}  
\usepackage{graphicx}
\usepackage{txfonts}
\usepackage{natbib}
\bibpunct{(}{)}{;}{a}{}{,}
\newcommand{\ros}{ROSAT}
\newcommand{\chan}{Chandra}
\newcommand{\xmm}{XMM-Newton}
\newcommand{\eROS}{eROSITA}
\newcommand{\agile}{AGILE}

\newcommand{\fermi}{Fermi-LAT}
\newcommand{\eso}{ESO-VLT}
\newcommand{\soar}{SOAR}
\newcommand{\lbt}{LBT}
\newcommand{\nh}{N_{\rm H}}
\newcommand{\etacar}{$\eta$\,Car}
\def \msev{M7}
\def \magoneeig{\object{RX~J1856.5-3754}}

\def \magonesix{\object{RX~J1605.3+3249}}
\def \magonethr{\object{RX~J1308.6+2127}}

\def \jtenfull{\object{2XMM~J104608.7-594306}}
\def \jten{J1046}
\def \rrat{\object{RRAT~J1819-1458}}
\def \calvfull{\object{1RXS~J141256.0+792204}}
\begin{document}
\title{The peculiar isolated neutron star in the Carina Nebula}
\subtitle{Deep \xmm\ and ESO-VLT observations of \jtenfull}
\author{A.~M.~Pires\inst{1}
    \and C.~Motch\inst{2}
    \and R.~Turolla\inst{3,4}
    \and A.~Schwope\inst{1}
    \and M.~Pilia\inst{5}
    \and A.~Treves\inst{6}
    \and S.~B.~Popov\inst{7}
    \and E.~Janot-Pacheco\inst{8}
    \fnmsep\thanks{Based on observations obtained with \xmm, an ESA science mission with instruments and contributions directly funded by ESA Member States and NASA (Target \jtenfull, \textsf{\small obsid}~0650840101). Optical observations were performed at the European Southern Observatory, Paranal, Chile, under programme IDs 382.D-0687(A) and 385.D-0209(A).}}
\offprints{A. M. Pires}
\institute{Leibniz-Institut f\"ur Astrophysik Potsdam (AIP), An der Sternwarte 16, 14482 Potsdam, Germany, 
    \email{apires@aip.de} 
    \and
    CNRS, Universit\'e de Strasbourg, Observatoire Astronomique, 11 rue de l'Universit\'e, F-67000 Strasbourg, France
    \and
    Universit\'a di Padova, Dipartimento di Fisica e Astronomia, via Marzolo 8, 35131 Padova, Italy
    \and 
    Mullard Space Science Laboratory, University College London, Holmbury St. Mary, Dorking, Surrey, RH5 6NT, UK
    \and
    Netherlands Foundation for Research in Astronomy, Postbus 2, 7990 AA, Dwingeloo, The Netherlands
    \and 
    Universit\'a dell'Insubria, Dipartimento di Fisica e Matematica, Via Valleggio 11, 22100 Como, Italy
    \and
    Sternberg Astronomical Institute, Lomonosov Moscow State University, Universitetskii pr. 13, 119991 Moscow, Russia
    \and
    Instituto de Astronomia, Geof\'isica e Ci\^encias Atmosf\'ericas, Universidade de S\~ao Paulo, \\ R. do Mat\~ao 1226, 05508-090 S\~ao Paulo, Brazil}
\date{Received ...; accepted ...}
\keywords{stars: neutron --
    X-rays: individuals: \jtenfull, \calvfull, \rrat}
\titlerunning{The peculiar isolated neutron star in Carina}
\authorrunning{A.~M.~Pires et al.}
\abstract{
While fewer in number than the dominant rotation-powered radio pulsar population, peculiar classes of isolated neutron stars (INSs) -- which include magnetars, the \ros-discovered ``Magnificent Seven'' (\msev), rotating radio transients (RRATs), and central compact objects in supernova remnants (CCOs) -- represent a key element in understanding the neutron star phenomenology.
We report the results of an observational campaign to study the properties of the source \jtenfull, a newly discovered thermally emitting INS.   
The evolutionary state of the neutron star is investigated by means of deep dedicated observations obtained with the \xmm\ Observatory, the ESO Very Large Telescope, as well as publicly available $\gamma$-ray data from the Fermi Space Telescope and the AGILE Mission. 
The observations confirm previous expectations and reveal a unique type of object. The source, which is likely within the Carina Nebula ($\nh=2.6\times10^{21}$\,cm$^{-2}$), has a spectrum that is both thermal and soft, with $kT_\infty=135$\,eV. Non-thermal (magnetospheric) emission is not detected down to 1\% ($3\sigma$, $0.1-12$\,keV) of the source luminosity. Significant deviations (absorption features) from a simple blackbody model are identified in the spectrum of the source around energies 0.6\,keV and 1.35\,keV. While the former deviation is likely related to a local oxygen overabundance in the Carina Nebula, the latter can only be accounted for by an additional spectral component, which is modelled as a Gaussian line in absorption with $\textrm{EW}=91$\,eV and $\sigma=0.14$\,keV ($1\sigma$). Furthermore, the optical counterpart is fainter than $m_{\rm V}=27$ ($2\sigma$) and no $\gamma$-ray emission is significantly detected by either the Fermi or \agile\ missions. 
Very interestingly, while these characteristics are remarkably similar to those of the \msev\ or the only RRAT so far detected in X-rays, which all have spin periods of a few seconds, we found intriguing evidence of very rapid rotation, $P=18.6$\,ms, at the $4\sigma$ confidence level. We interpret these new results in the light of the observed properties of the currently known neutron star population, in particular those of standard rotation-powered pulsars, recycled objects, and CCOs. 
We find that none of these scenarios can satisfactorily explain the collective properties of \jtenfull, although it may be related to the still poorly known class of Galactic anti-magnetars. Future \xmm\ data, granted for the next cycle of observations (AO11), will help us to improve our current observational interpretation of the source, enabling us to significantly constrain the rate of pulsar spin down.
}
\maketitle
\section{Introduction\label{sec_intro}}
A major outcome of the \ros\ mission was the discovery of a group of seven radio-quiet thermally emitting isolated neutron stars (INSs), which were originally identified serendipitously as soft, bright X-ray sources with no obvious optical counterparts. They share a rather well-defined set of properties that have never been encountered together in previously known classes of INSs, and have been nicknamed the ``Magnificent Seven'' (\msev, for short; see \citealt{hab07,kap08a,tur09}, for reviews). 

Relative to standard (rotation-powered) radio pulsars, these INSs rotate slower ($P\sim10$\,s), have thermal X-ray luminosities higher than their spin-down power, and stronger magnetic fields ($B\gtrsim10^{13}$\,G, as inferred from timing measurements as well as from the broad absorption lines in their X-ray spectra; see \citealt[][and references therein]{kap11b}). The \msev\ display neither persistent nor transient radio emission to a rather sensitive limiting flux \citep{kon09} and are unassociated with supernova remnants. Proper motion studies \citep[][and references therein]{mot09} show that they are cooling, middle-aged neutron stars (ages $10^5$ to $10^6$\,yr), probably born in the nearby OB associations of the Gould Belt. Their proximity (few hundred parsecs; \citealt{pos07,ker07a}) and the combination of strong thermal radiation and absence of significant magnetospheric activity make them ideal targets for testing surface emission models, deriving radii and constraining the equation of state of neutron star matter. Unfortunately, the current lack of understanding of the surface composition, magnetic field, and temperature distributions have limited any definite conclusion so far \citep[see e.g.][]{kap11a}.

Growing evidence relates the \msev\ to other peculiar groups of INSs, in particular the magnetar candidates, anomalous X-ray pulsars, and soft $\gamma$-ray repeaters (AXPs and SGRs, see e.g. \citealt{mer08}, for a review), and the rotating radio transients (RRATs, \citealt{lau06,kea11b}).
The long spin periods, bright thermal emission, and intense magnetic fields suggest that some of the \msev\ might have evolved from the younger and more energetic magnetar objects \citep[e.g.][]{hey98,pop10}. Crustal heating by means of magnetic field decay seems to play an important role in the thermal evolution of neutron stars with $B\gtrsim5\times10^{13}$\,G, since a correlation between $B$ and $kT$ is observed \citep[e.g.][]{agu08,pon09,kap09d}. 
On the other hand, the discovery of RRATs is intriguing since these sources have so far manifested themselves in a variety of ways. While many, in spite of their transient nature, display timing properties that cannot be distinguished from those of the bulk of normal radio pulsars, several RRATs with longer spin periods and higher magnetic fields occupy a region of the $P-\dot{P}$ diagram that is devoid of normal pulsars and close to that populated by the \msev.
Moreover, the most active source among the known sample is the highly magnetized \rrat, the only one that has so far been detected in X-rays \citep{rey06,lau07,kap09c}\footnote{Of the more than 60 RRATs known to date, $\sim20$ have precise position determinations and six have been investigated in X-rays; see \citet{kea11a} and \citet{kea11b} for reviews.}. Very interestingly, the X-ray source was found to exhibit a spectrum that although hotter and more absorbed is remarkably similar to those of the \msev. Unusual timing behaviour following glitches detected in the radio indicates that this source might also have evolved from a magnetar \citep{lyn09}. 

It is remarkable that a group of very similar sources, displaying at the same time unique properties that are extremely different from those of standard radio pulsars, are all detected in the very local solar vicinity. Taking into account that the \msev\ represent about half of all young (younger than $3$\,Myr) INSs known within $\sim$\,$1$\,kpc \citep{pop03}, they could be just the tip of an iceberg of a largely hidden population of stellar remnants.
Both the poorly constrained populations of RRATs and thermally emitting INSs are indeed estimated at present to outnumber ordinary radio pulsars by factors of one to three \citep[e.g.][]{kea10a}, whereas the latter alone are sufficient to account for the total number of past Galactic core-collapse supernovae. Therefore, to avoid ``superpopulating'' the Galaxy with unrelated neutron stars, it is necessary to invoke links between the several subgroups, which may be either selection effects (i.e. the same object appears differently to the observer due to viewing biases) or actual evolutionary relations. 
\subsection{The source \jtenfull\label{sec_J1046}}
Radio-quiet INSs are extremely elusive and difficult to detect. For several years, considerable efforts have been devoted to discovering new sources similar to the \msev\ \citep[e.g.][]{rut03,chi05,agu06,tur10,agu11}. These involve cross-correlating X-ray sources -- often with large positional errors -- with a large number of catalogued objects at other wavelengths, and dealing appropriately with the level of sample contamination by other classes of X-ray emitters. In spite of these searches, only two radio-quiet thermally emitting INSs have been positively identified to date since the \ros\ era, namely Calvera (\calvfull) and \jtenfull\ (hereafter \jten). 

Calvera is a relatively bright \ros\ source, which was originally identified as a likely compact object based on its large X-ray-to-optical flux ratio \citep{rut08}. The distance to the source is poorly constrained. Its interpretation as a cooling INS (with a similar size of emission radius as the \msev) has always met with difficulties, since its high Galactic latitude and hot temperature ($kT\sim200$\,eV) would either require a very high spatial velocity to explain its current position well above the Galactic plane or point to a non-standard cooling or re-heating process. Alternatively, if Calvera's progenitor were a high-velocity runaway star, it might have been born away from the Galactic plane \citep{pos08}.
Recent multiwavelength investigations by \citet{zan11} confirmed the radio-quiet nature of the source and most notably revealed the neutron star's very short spin period of $P\sim59$\,ms, which was later also apparently detected in $\gamma$-rays with data from the Fermi Large Area Telescope \citep[LAT,][]{atw09}. The nature of Calvera however remains unclear, especially after \citet{hal11} called the $\gamma$-ray detection into question, leaving the spin-down rate of the source far more poorly constrained (see Sect.~\ref{sec_disc}).

We previously reported the results of a programme to identify new thermally emitting INSs in the \xmm\ catalogue of serendipitous X-ray sources \citep{pir09b}. 
The X-ray brightest INS candidates resulting from our search have been the scope of succesful proposals carried out over the past four years in the optical with the 8.1\,m European Southern Observatory Very Large Telescope (\eso), the two 8.4\,m Large Binocular Telescope (\lbt), and the 4.1\,m Southern Observatory for Astrophysical Research (\soar). These deep optical investigations permitted the discovery of a new thermally emitting INS, \jten\ \citep[][hereafter Paper~I]{pir09a}\defcitealias{pir09a}{Paper~I}. 

As expected for a cooling INS, \jten\ is characterised by a soft blackbody-like emission, stable observed flux over a long timescale, a very high X-ray-to-optical flux ratio, and no radio counterpart in the Parkes Multibeam Pulsar Survey (which has a sensitivity of 0.14\,mJy for a canonical pulsar). The column density is consistent with the source being located within the Carina Nebula, at a distance of 2\,kpc \citepalias{pir09a}. This giant \ion{H}{II} region is indeed likely to harbour unidentified cooling neutron stars, which are the faint analogues of those produced in other close-by regions of intensive star formation \citep[][see their Fig.~7]{pos08}.

To investigate its properties and evolutionary state, we have ongoing dedicated projects in the optical and X-rays to observe this source. We report here the results of an observational campaign with \xmm\ and \eso.
The paper is structured as follows: in Sects.~\ref{sec_xrays} and \ref{sec_opt}, we describe the X-ray and optical follow-up observations, together with detailed analysis and results. A search for pulsations in \fermi\ and \agile\ \citep{tav09} data is described in Sect.~\ref{sec_gamma}. The summary and discussion of the results are in Sect.~\ref{sec_disc}. Finally, our conclusions are given in Sect.~\ref{sec_summary}.
\begin{table}[t]
\caption{Instrumental configuration and duration of the EPIC scientific exposures of the \xmm\ observation of \jtenfull
\label{tab_exposureinfoEPIC}}
\centering
\begin{tabular}{l c c r}
\hline\hline
Instr. & Start Time & Mode & Durat. \\
       & (UTC)      &      & (s)      \\
\hline
pn &	2010-12-06T00:14:10.0  & Prime small window &  90,471\\ 
MOS1 &	2010-12-06T10:01:39.0  & Prime partial W2   &  55,120\\ 
MOS2 &	2010-12-06T10:01:45.0  & Prime partial W2   &  55,129\\ 
\hline
\end{tabular}
\tablefoot{The cameras were operated in imaging mode and the thin filter was used (\textsf{\small obsid}~0650840101).}
\end{table}
\section{\xmm\ observations\label{sec_xrays}}
Being located at an angular distance of $\sim8.5'$ from the well-studied binary system Eta Carinae (\etacar), \jten\ has been detected on many occasions by \xmm\ and \chan. Unfortunately, in most observations, conditions were far from optimal, given the source's large off-axis angles, $\sim9'$ on average, and short effective exposure times, usually less than 15\,ks \citepalias[see][for details]{pir09a}\footnote{In spite of the ten-year-long timebase, a proper motion study in X-rays is impractical due to the source's estimated distance ($d\sim2$\,kpc) and to the large off-axis angles and corresponding positional errors in the observations before AO9.}. We were granted \xmm\ time (90\,ks, AO9) to help us to characterize the source spectral energy distribution and search for pulsations. 
\subsection{Observations and data reduction\label{sec_xmmobs_datareduction}}
Observations were carried out on 2010 December 6, for a total exposure time of 90.917\,ks. All instruments on-board \xmm\ \citep{jan01} were active during the observation. Table~\ref{tab_exposureinfoEPIC} contains information on the scientific exposures and instrumental setup of the EPIC pn \citep{str01} and MOS \citep{tur01} detectors.

The EPIC cameras operated in small window (SW) mode with thin filters. In this configuration, pn and MOS provide a time resolution of 6\,ms and 0.3\,s, respectively; owing to the more frequent readout, the nominal exposures are shorter by factors of 0.71 and 0.975 (corresponding to deadtimes of 29\% and 2.5\%). For pn, only the $63\times64$ pixels ($4.3'\times4.3'$) of CCD 4 are read out. For MOS, the central $100\times100$ pixel area ($1.8'\times1.8'$) of the inner CCD operates with a higher time resolution, while the outer chips are read out normally.

The MOS cameras experienced sporadic \textit{full scientific buffers} during the first 35\,ks of the observation, alternating very short scientific exposures, of typical durations of shorter than half a kilosecond, with periods where the cameras operated in the so-called ``counting mode'', lasting between 2.6\,ks and 6.8\,ks. In this mode, no transmission of information for individual X-ray events occurs.
As a result, the effective observing time is considerably shorter for MOS. 
The reason for the abnormal operation of the MOS detectors is the presence of the X-ray bright \etacar\ in one of the peripheral MOS CCDs, which are used to monitor the radiation level during the observation under the assumption that no bright sources are present. The MOS exposures were unfortunately interrupted and restarted several times until the origin of the high count-rate level could be identified (R.~Gonzalez-Riestra, private communication).

Data reduction was performed with \textsf{\small SAS~11} (\textsf{\small xmmsas-20110223-1801}) by applying standard procedure and using the latest calibration files.
MOS exposures U008 and pn S003 were processed using the EPIC meta tasks \textsf{\small emchain} and \textsf{\small epchain}, respectively, applying default corrections. We ensured that the pn event files were clean of unrecognised time jumps (i.e. those uncorrected by standard \textsf{\small SAS} processing).
\begin{table}[t]
\caption{Source coordinates with $1\sigma$ errors
\label{tab_coordinates}}
\centering
\begin{tabular}{l c c c}
\hline\hline
Camera & RA & DEC & $\sigma$\\
       & (h m s) & (d m s) & (arcsec)    \\
\hline
pn     & 10\ \ 46\ \ 08.500 & -59\ \ 43\ \ 05.179 & 0.2 \\
MOS1   & 10\ \ 46\ \ 08.536 & -59\ \ 43\ \ 04.863 & 0.3 \\
MOS2   & 10\ \ 46\ \ 08.485 & -59\ \ 43\ \ 04.451 & 0.3 \\
EPIC   & 10\ \ 46\ \ 08.503 & -59\ \ 43\ \ 05.143 & 0.2 \\
\hline
EPIC$^\dag$ & 10\ \ 46\ \ 08.756 & -59\ \ 43\ \ 05.523 & 0.4 \\
\hline
\chan  & 10\ \ 46\ \ 08.719 & -59\ \ 43\ \ 06.480 & $<1$\\
\xmm   & 10\ \ 46\ \ 08.730 & -59\ \ 43\ \ 06.300 & 0.4 \\
\hline
\end{tabular}
\tablefoot{We also list the position of the source as previously determined in archival \chan\ and \xmm\ data. $^\dag$Corrected position after cross-correlating the EPIC source list with optical objects in the field (see text).}
\end{table}

No background flares were registered during the 90\,ks (last 55\,ks) of the pn (MOS) observations. The effective observing times were therefore only reduced by the cameras' livetime in SW mode: 63.25\,ks for pn and 53.4\,ks for MOS. We note that although optical loading (due to unblocked optical photons from \etacar\ and other bright OB stars in the field-of-view) is obviously present at \etacar's position in the MOS images, out-of-time events or charge transfer efficiency (CTE) alteration are unimportant at the position of the target at aimpoint, since \etacar\ is not located on the same CCD as \jten\ or along the readout direction (M.~Ehle, private communication).

For the analysis, we filtered the event lists to retain the pre-defined photon patterns with the highest quality energy calibration -- single and double events for pn (pattern $\le4$) and single, double, triple, and quadruple for MOS (pattern $\le12$) -- as well as to exclude bad CCD pixels and columns. Source photons were extracted from circular regions of radius $18''$ (unless otherwise noted) centred on the position of the source (as given by \textsf{\small emldetect} for each camera; see Sect.~\ref{sec_srcdet}). Background circular regions of size $50''$ (pn) and $25''$ (MOS) were defined off-source, on the same CCD as the target. 
Owing to the frequent readout of the pn in SW mode, detector noise dominates the low-energy count distribution. We therefore restricted the analysis of pn data to photons with energies between 0.3\,keV and 2\,keV (see however Sect.~\ref{sec_timing}). For MOS, the energy band for the analysis is $0.15-2$\,keV.

The detected source count rates are $7.74(14)$, $1.66(6)$, and $1.90(6)$ ($\times10^{-2}$\,s$^{-1}$; pn, MOS1, and MOS2); these closely agree with the expectations of simulations (considering the source parameters as in \citetalias{pir09a}). Using the \textsf{\small SAS} task \textsf{\small epiclccorr}, we created lightcurves for \jten, which were corrected for bad pixels, deadtime, exposure, and background counts. The statistics for these corrected lightcurves, binned into 900\,s and 550\,s (pn and MOS) intervals, show the $3\sigma$ upper limits for the rms fractional variation of 0.12, 0.30, and 0.25 for pn, MOS1, and MOS2, respectively. 
On the basis of the same lightcurves, the reduced $\chi^2_\nu$ assuming a constant flux is 1.007 (pn, 100 d.o.f.), 0.979, and 1.067 (MOS1 and MOS2; 99 d.o.f.), corresponding to null-hypothesis probabilities of 46\%, 51\%, and 28\%. 
\begin{table*}[t]
\caption{Parameters of \jtenfull, as extracted from the \xmm\ AO9 observations
\label{tab_sourceMLparam}}
\centering
\begin{tabular}{l l l l l}
\hline\hline
Parameter                       & pn                     & MOS1                     & MOS2                     & EPIC \\
\hline
Counts                          & $5594\pm91$            & $1024\pm41$            & $1133\pm40$            & $7912\pm108$ \\
Counts ($0.2-0.5$\,keV)         & $1236\pm41$            & $154\pm15$             & $161\pm15$             & $1258\pm42$\\
Counts ($0.5-1.0$\,keV)         & $3437\pm71$            & $611\pm32$             & $627\pm29$             & $3555\pm71$\\
Counts ($1.0-2.0$\,keV)         & $910\pm37$             & $257\pm21$             & $343\pm22$             & $921\pm37$\\
Counts ($2.0-4.5$\,keV)         & $0\pm7$                & $2\pm4$                & $1.4\pm2.5$            & $2\pm5$\\
Counts ($4.5-12$\,keV)          & $11\pm12$              & $0\pm4$                & $0\pm5$                & $14\pm12$\\
Detection likelihood            & $6522$                 & $968$                  & $1519$                 & $9436$\\
Flux ($\times10^{-13}$\,erg\,s$^{-1}$\,cm$^{-2}$)& $1.17(4)$ & $1.09(7)$          & $1.17(7)$              & $1.174(29)$\\
Rate ($\times10^{-2}$\,s$^{-1}$)& $9.03(15)$             & $2.00(8)$              & $2.17(8)$              & $13.43(18)$\\
$l$ (degrees)                   & $287.73$               & $287.73$               & $287.73$               & $287.73$\\
$b$ (degrees)                   & $-0.59591$             & $-0.59579$             & $-0.59574$             & $-0.59587$\\
HR$_1$                          & $+0.471\pm0.015$       & $+0.60\pm0.04$         & $+0.59\pm0.03$         & $+0.477\pm0.015$\\
HR$_2$                          & $-0.581\pm0.015$       & $-0.41\pm0.04$         & $-0.29\pm0.04$         & $-0.588\pm0.015$\\
HR$_3$                          & $-1.000\pm0.016$       & $-0.985\pm0.028$       & $-0.992\pm0.014$       & $-0.997\pm0.011$\\   
\hline
\end{tabular}
\tablefoot{Counts, fluxes, and rates are given in the total \xmm\ energy band ($0.2-12$\,keV).}
\end{table*}
\subsection{Source detection\label{sec_srcdet}}
To measure the position of \jten, we created images in the five \xmm\ pre-defined energy bands, for each EPIC camera, and analysed them (individually and simultaneously) with the source-detection meta task \textsf{\small edetect\_chain}. 
In Table~\ref{tab_coordinates}, we list the source equatorial coordinates in each camera together with previous determinations based on \chan\ and \xmm\ (\textsf{\small obsid}~0311990101, with the highest signal-to-noise ratio) archival data. 

The position of the source was found to be offset relative to previous determinations. By cross-correlating the list of (pipeline-processed) EPIC X-ray sources with those of optical (GSC2.3) objects lying within $17'$ from \jten, we found offsets of $-1.8''$ in RA and $0.9''$ in DEC, as well as a slight rotational offset of $0.013^\circ$, using the \textsf{\small SAS} task \textsf{\small eposcorr} and a number of 59 matches. The corrected source position, updated accordingly, is consistent with previous determinations (Table~\ref{tab_coordinates}). We calculated the $1\sigma$ errors in the updated position by quadratically adding a systematic uncertainty of $0.35''$ to the nominal errors computed by \textsf{\small emldetect}, which is the procedure adopted by the catalogue pipeline of \xmm, based on correlations with optical catalogues.
\begin{figure*}
\begin{center}
\includegraphics*[width=0.495\textwidth]{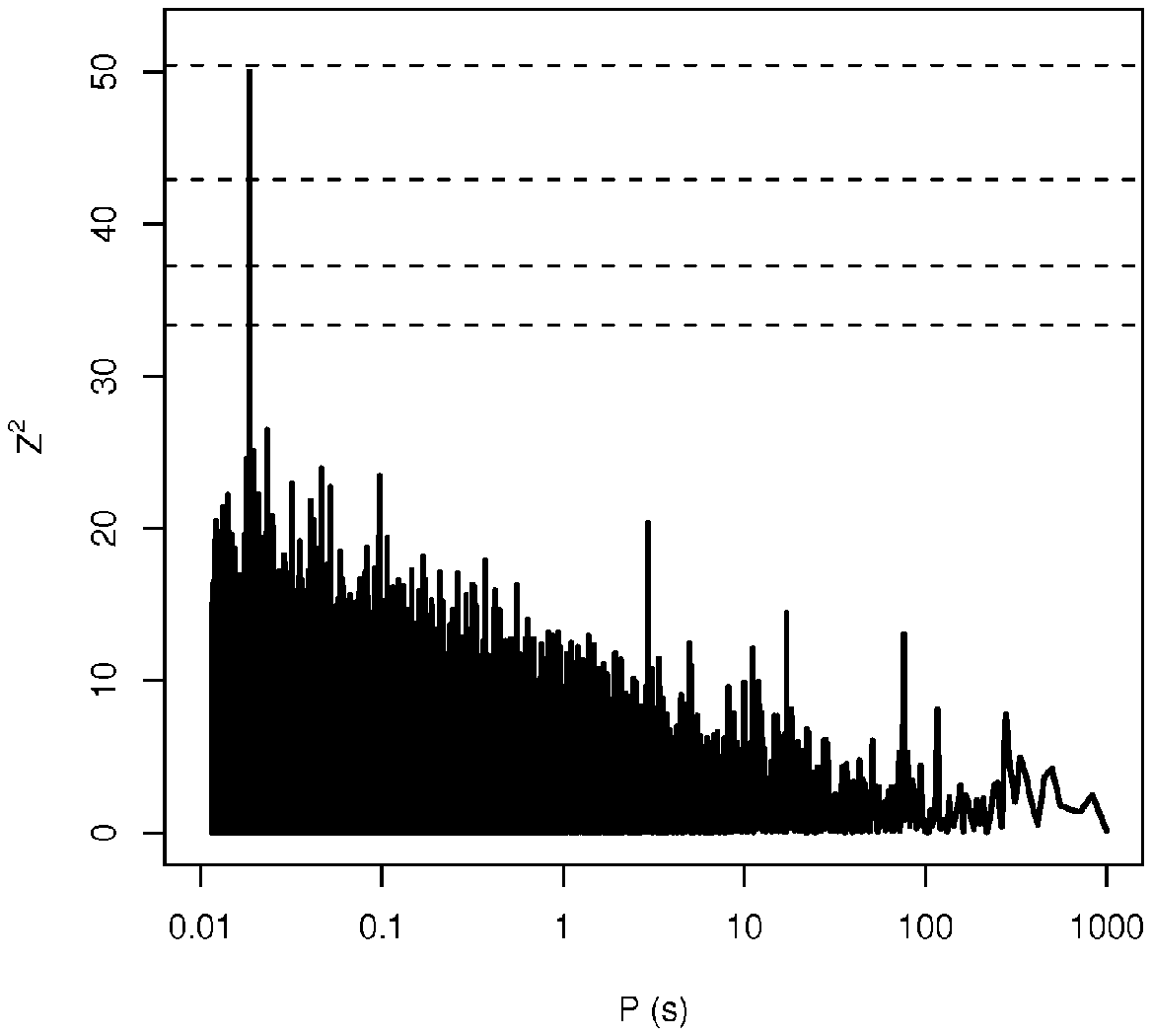}
\includegraphics*[width=0.495\textwidth]{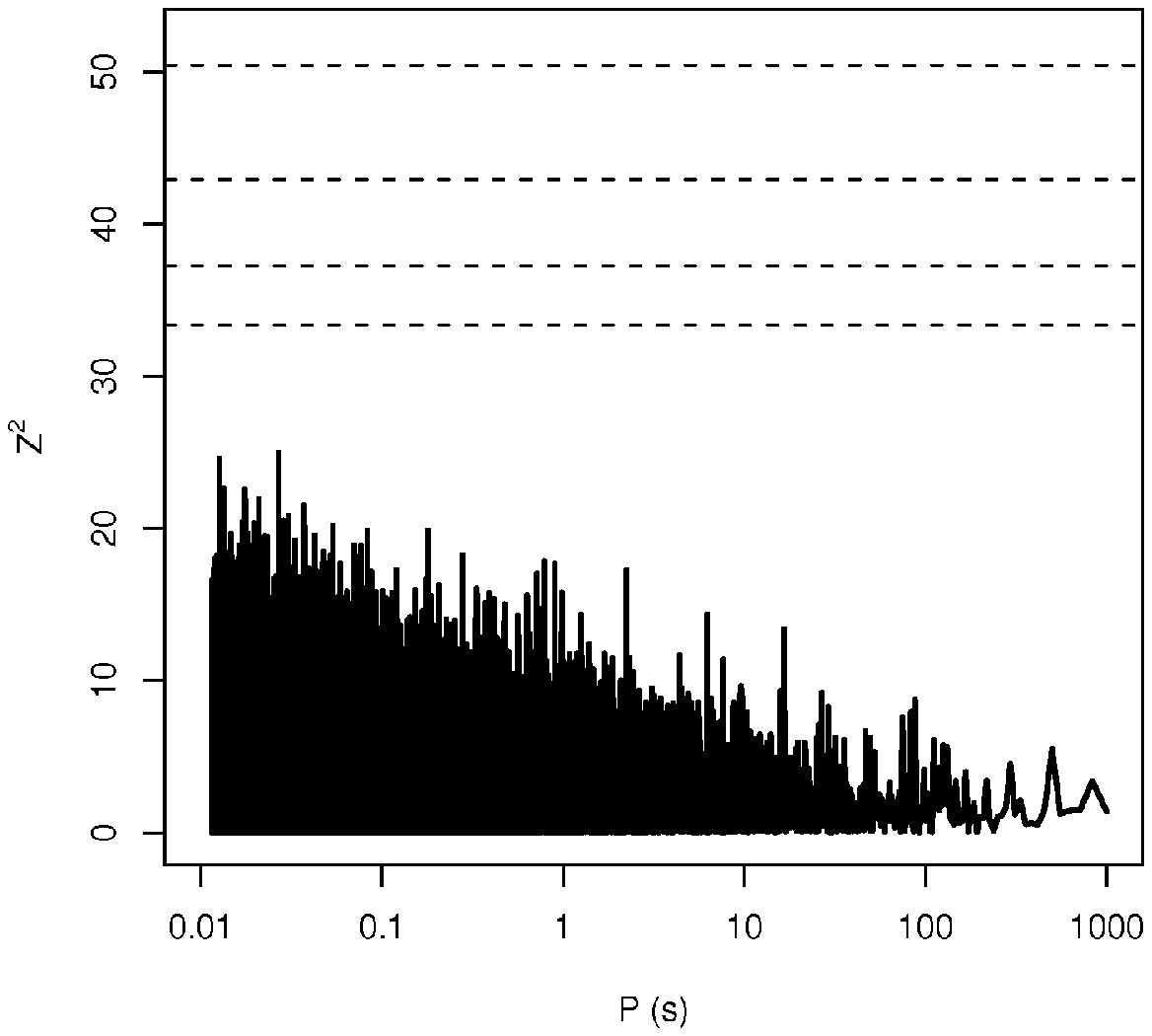}
\end{center}
\caption{Results of the $Z^2_1$ search (pn data, $P=0.011-1000$\,s). The frequency range is $\Delta\nu=87.72$\,Hz, the energy band is $0.36-2.2$\,keV, and the size of the extraction region is $18.85''$ (5260 counts). Dashed horizontal lines show confidence levels of from $1\sigma$ to $4\sigma$ for the detection of a periodic signal, given the frequency range and length of the observation. A periodic signal at $P_\ast\sim18.6$\,ms is detected at $4\sigma$ (left). The periodogram on the right shows the same analysis conducted for a background region (size $52''$ and 5322 total counts; see text).}\label{fig_z2best}
\end{figure*}

Source parameters based on a maximum likelihood fitting are given in Table~\ref{tab_sourceMLparam}. The hardness ratios and corresponding errors were computed according to

\begin{displaymath}
\left\{ \begin{array}{llll}
\mathrm{HR}_{i} & = & \frac{\displaystyle C_{i+1}-C_{i}}{\displaystyle C_{i+1}+C_{i}} &\qquad i=1,\ldots,4,\\
 & & & \\
\sigma_{\mathrm{HR}_i} & = & \frac{\displaystyle 2}{\displaystyle (C_{i+1}+C_{i})^2}\sqrt{C_{i}^2\sigma_{C_{i+1}}^2 + C_{i+1}^2\sigma_{C_{i}}^2}, 
\end{array} \right.
\end{displaymath}
\noindent where $C_i$ are counts in a given energy band and $\sigma_{C_{i}}$ are the corresponding errors. 
Owing to the low counts detected in energy bands 4 and 5 (above 2\,keV), the errors in $\mathrm{HR}_4$ are too large; hence we only list in Table~\ref{tab_sourceMLparam} the values of hardness ratios that can be constrained by the errors ($\mathrm{HR}_{1,2,3}$).
\subsection{Timing analysis\label{sec_timing}}
Previous X-ray observations of \jten\ did not reveal pulsations to a poorly constraining 30\% upper limit ($3\sigma$ confidence level), in the $0.15-100$\,s period range \citepalias{pir09a}. 
Simulations showed that for an exposure time of 90\,ks the EPIC cameras operating with the thin filter can collect enough photons to significantly ($>4\sigma$) detect pulsations from \jten\ of periods in the range $P=0.6-1000$\,s, for pulsed fractions higher than $\sim10-15$\%.
The time resolution of the EPIC cameras in SW mode allows us to search for pulsations with frequencies as high as 1.7\,Hz -- similar to those observed in both RRATs and the \msev. The higher resolution of pn in SW mode also allows us to search for higher frequencies, up to 88\,Hz, albeit with lower sensitivity.
\begin{figure*}
\begin{center}
\includegraphics*[width=0.495\textwidth]{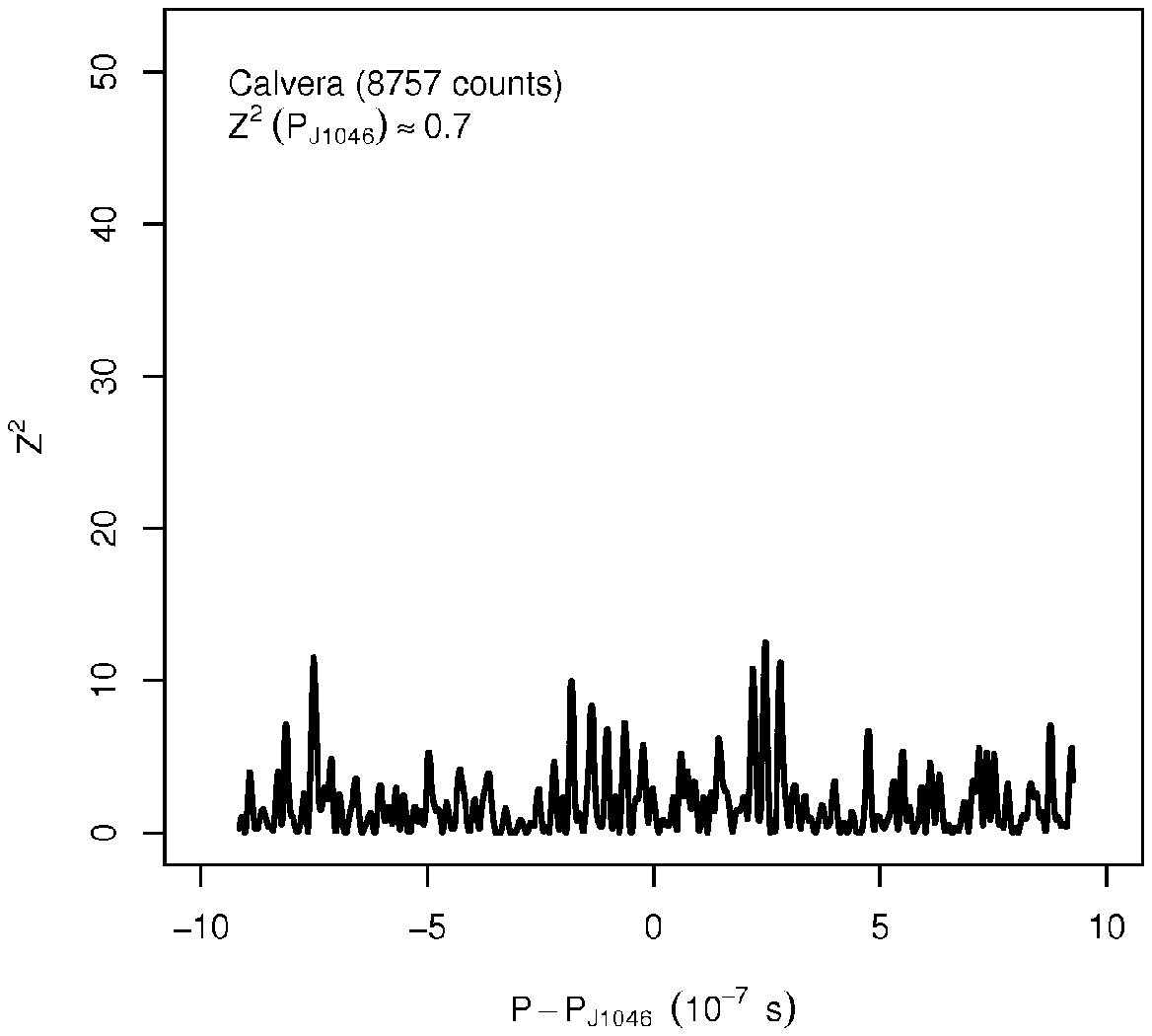}
\includegraphics*[width=0.495\textwidth]{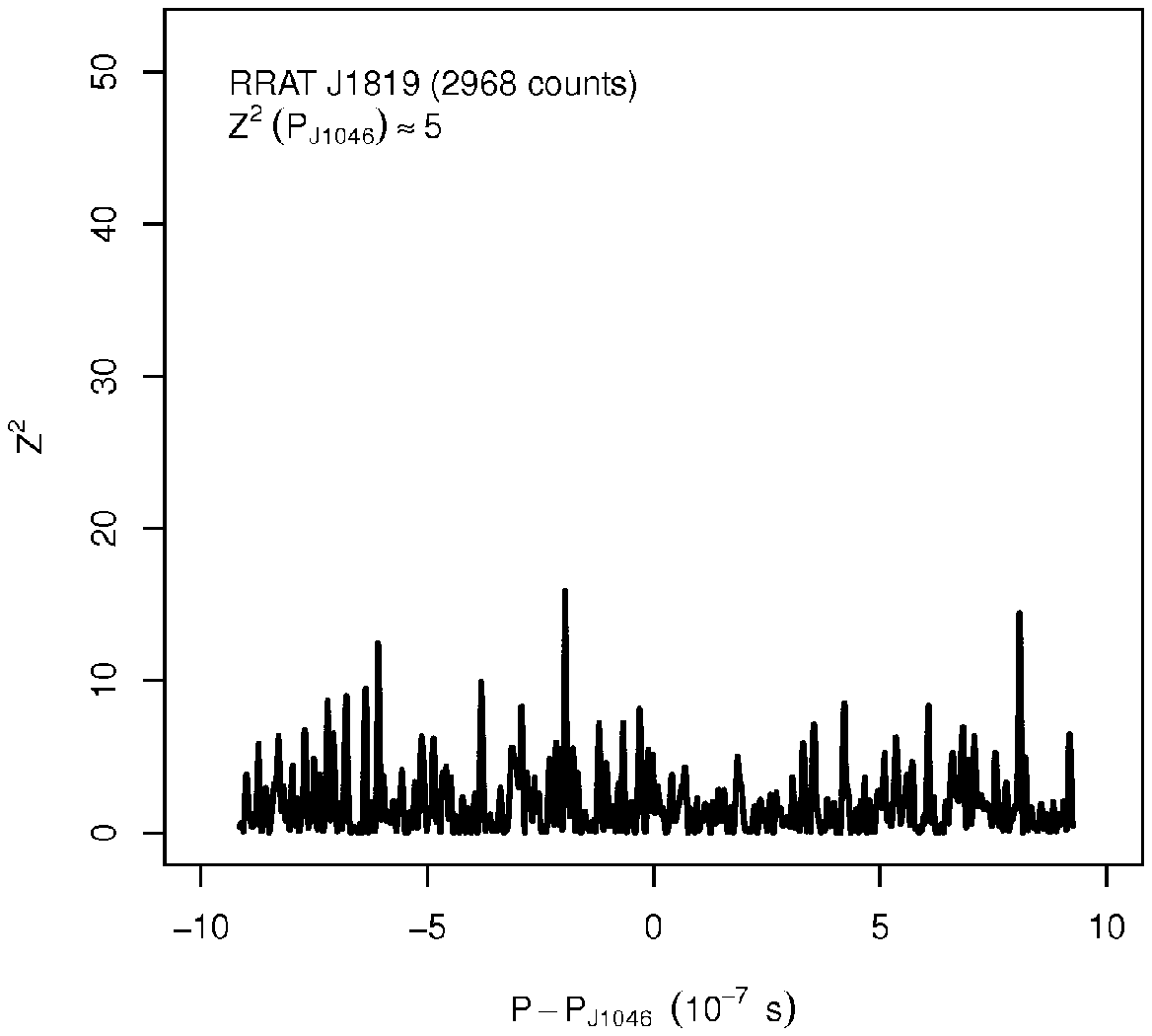}
\includegraphics*[width=0.495\textwidth]{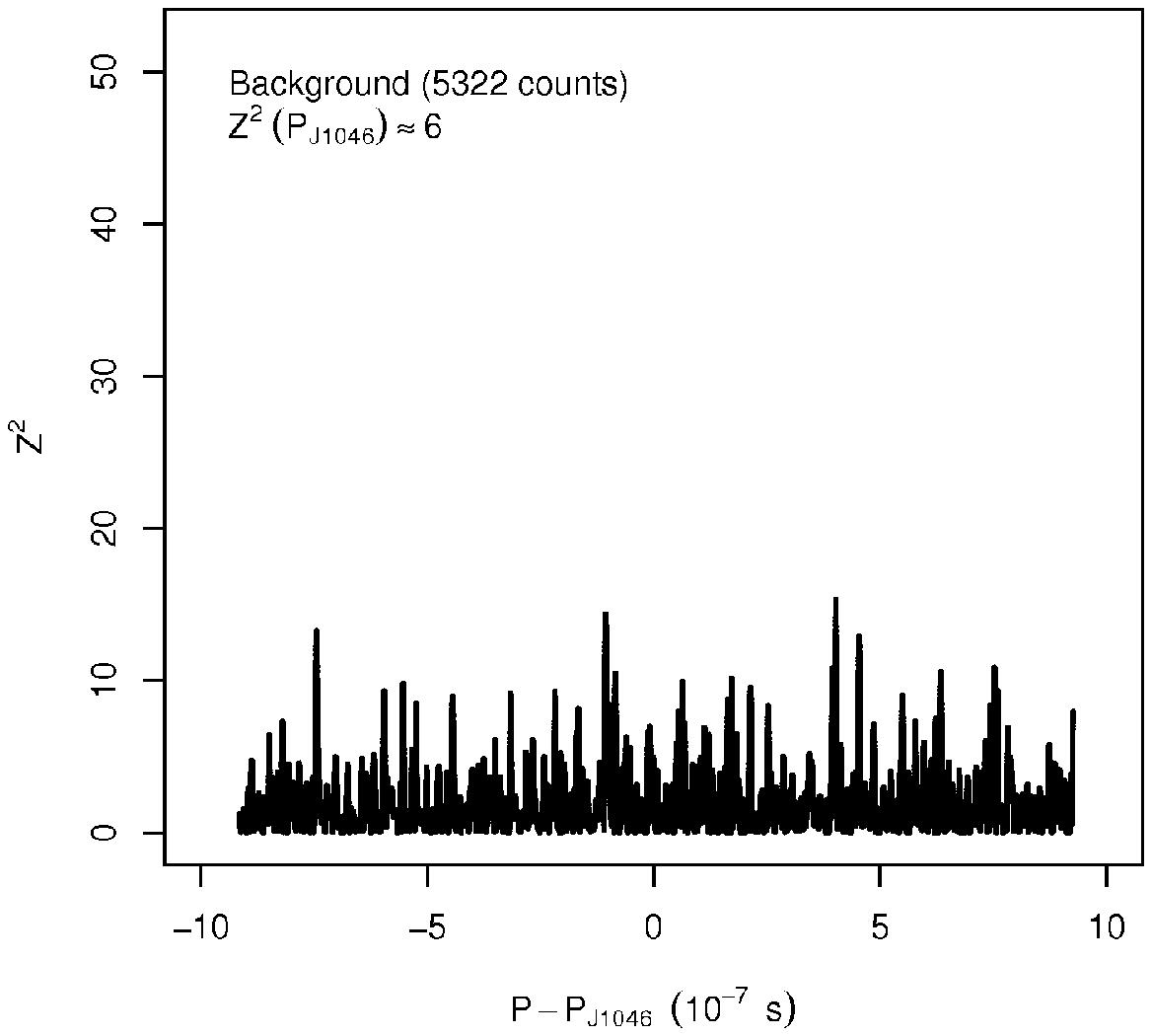}
\includegraphics*[width=0.495\textwidth]{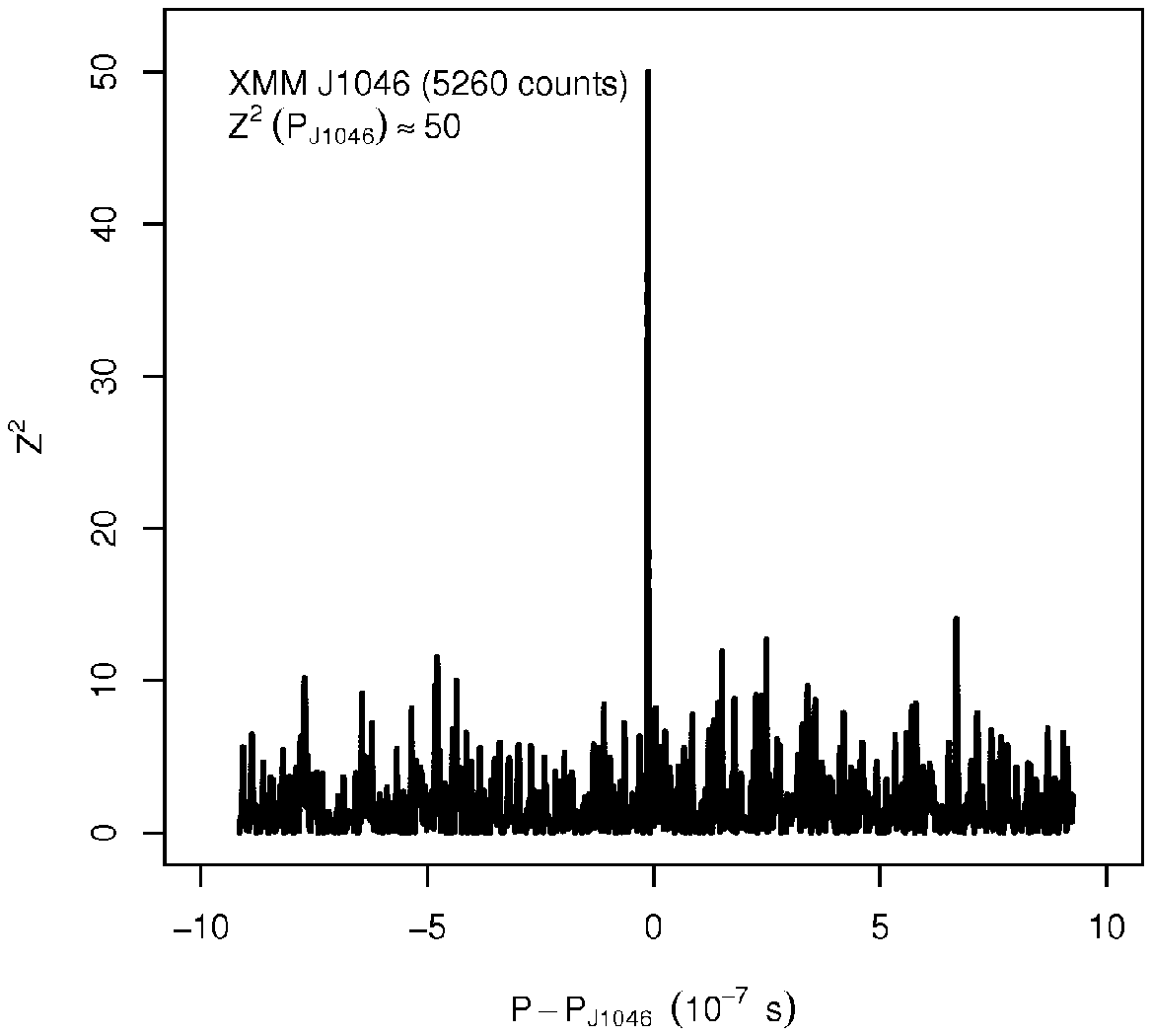}
\end{center}
\caption{Our $Z^2_1$ analysis around the periodicity at $P_\ast\sim18.6$\,ms. The frequency range is $5.3\times10^{-2}$\,Hz and the step in frequency is 0.1\,$\mu$Hz; the energy band is $0.36-2.2$\,keV and the size of the extraction region is $18.85''$. We present the results of the same analysis for different data sets observed in SW mode: at top left, we show the periodogram for Calvera (\textsf{\small obsid}~0601180201; 8757 total counts); at top right, we show that for \rrat\ (\textsf{\small obsid}~0406450201; 2968 total counts). Below are the periodograms for our data: at the left we provide results for events extracted from a background region (5322 total counts for an extraction region of size $52''$) and on the right, for \jten\ (total of 5260 counts).}\label{fig_z2otherins}
\end{figure*}

For the timing analysis, we considered events with pattern 12 or lower. The times-of-arrival of the pn/MOS photons were converted from the local satellite to the solar system barycentric frame using the \textsf{\small SAS} task \textsf{\small barycen} and the source coordinates (Table~\ref{tab_sourceMLparam}).
A $Z^2_1$ (Rayleigh) test \citep{buc83} was used, which is appropriate when searching for smooth pulsations similar to those in the X-ray emission of thermally emitting INSs (in particular, four of the \msev\ show pulsed fractions, $p_{\rm f}$, lower than 10\%). 

We analysed the EPIC cameras together in the $P=0.6-10000$\,s period range, by testing different energy bands and extraction region radii. 
Given the high noise of the pn camera at low energies and the energy-dependent signal-to-noise ratio ($S/N$), we performed searches testing energy intervals where we aimed to achieve the optimal compromise between sensitivity and source signal over background counts. More specifically, we defined energy bands where the $S/N$ was both the highest, $S/N>5$ ($0.4-1.3$\,keV, $\sim4260-5890$ pn events), and moderate, $S/N>3$ ($0.33-1.5$\,keV, $\sim4640-6500$ pn events); we also extensively tested other energy configurations by varying the lower ($\epsilon_l$) and upper ($\epsilon_u$) ends of the energy band, $\epsilon_l$ between 0.2\,keV and 1\,keV and $\epsilon_u$ between 1.9\,keV and 2.5\,keV. 
The radii of the extraction regions were varied between $10''$ and $25''$ around the position of the source, as determined for each camera with \textsf{\small emldetect} (Table~\ref{tab_coordinates}).
To search for higher frequency pulsations, with periods as short as $P=0.0114$\,s, we restricted the analysis to the pn data, and tested different parameters (energy bands, extraction radii, and photon patterns).
\begin{figure*}
\begin{center}
\includegraphics*[width=0.495\textwidth]{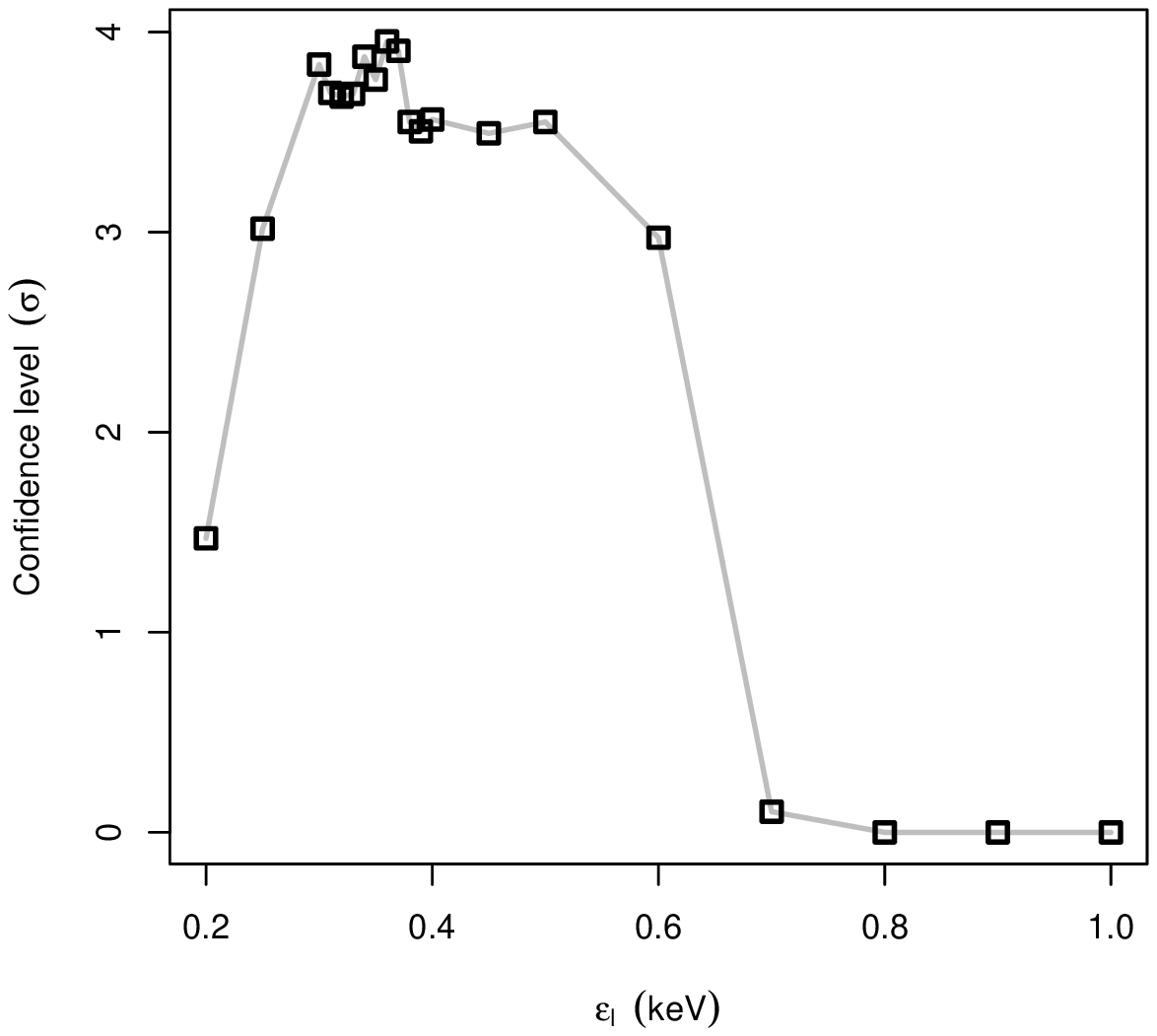}
\includegraphics*[width=0.495\textwidth]{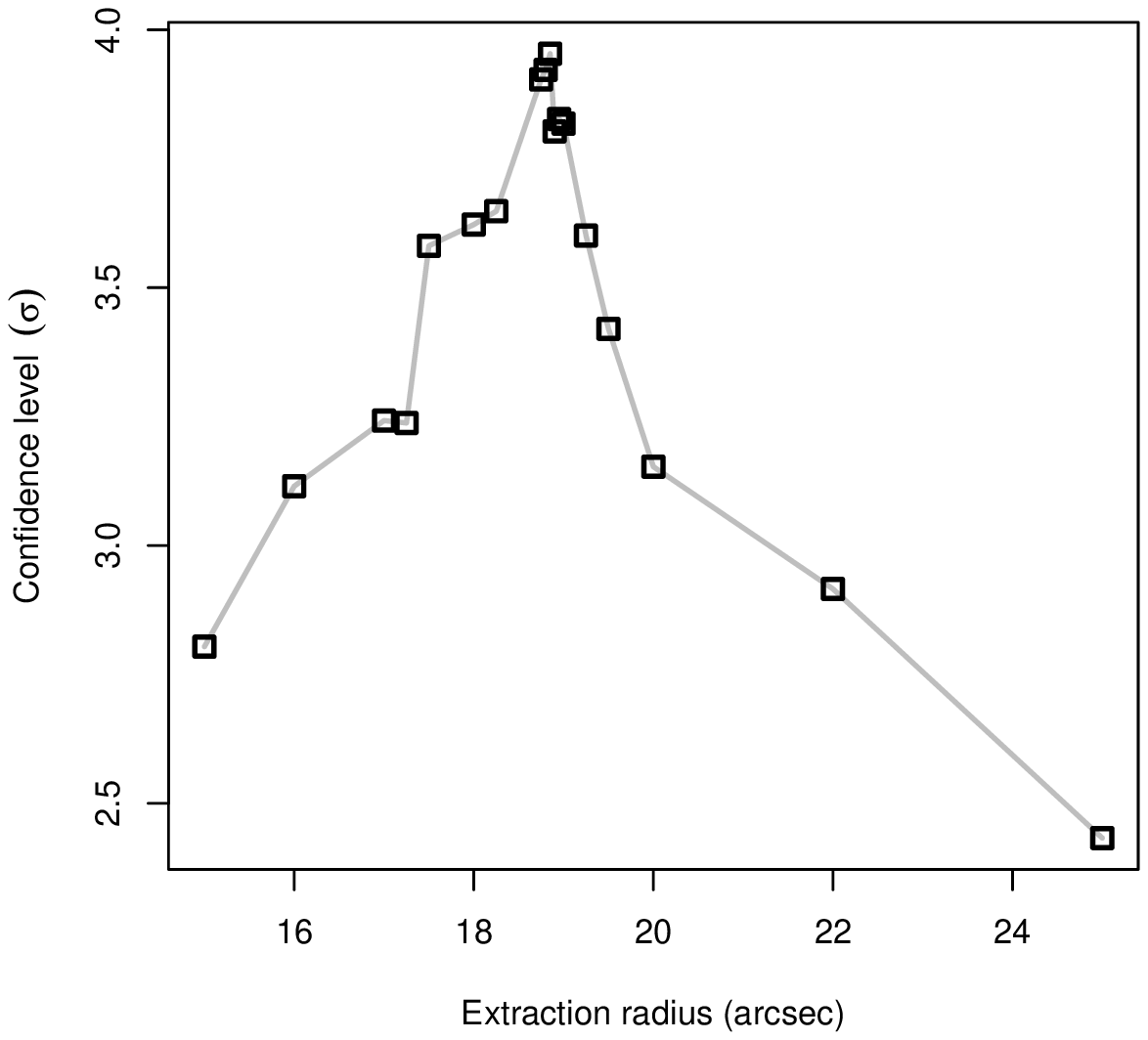}
\end{center}
\caption{\textit{Left.} Significance of the $Z^2_1$ statistics at $P=P_\ast\sim18.6$\,ms, as a function of the lower end of the energy band, $\epsilon_l-2.2$\,keV. The size of the extraction region is $18.85''$. \textit{Right.} Significance of the $Z^2_\ast$ statistics as a function of the size of the extraction region, for energy band $0.36-2.2$\,keV. The confidence levels are given relative to the entire frequency range of the timing analysis.}\label{fig_z2radband}
\end{figure*}
The adopted step in frequency was 1\,$\mu$Hz when analysing EPIC and twice this value when running the test on pn, typically for a frequency range $\sim50$ times broader. These values oversample the expected width of the $Z^2_1$ peaks by a factor of nearly ten, warranting that a peak corresponding to a periodic signal is not missed. The number of statistically independent trials depends on the chosen frequency range of the search and the exposure time of the observation; typically it is $\sim10^5$ for EPIC searches and between $\sim10^5$ and $6\times10^6$ when screening pn data. 

In the frequency range where we are most sensitive to pulsations, namely periods $P=0.6-10000$\,s (when the EPIC cameras are analysed together), no significant periodicities were found. 
The $3\sigma$ upper limit to the source pulsed fraction in this period range is 9.6\%, considering the $0.3-2.5$\,keV energy band. 

Very interestingly, extending the search to higher frequencies (pn data only), we found evidence of a periodicity at $P\equiv P_\ast=18.640787(6)$\,ms. The peak at frequency $\nu_\ast=53.645804(16)$\,Hz has a maximum power of $Z^2_\ast\sim50$ and corresponds to a detection at $4\sigma$ (see below). The resulting periodogram can be seen in Fig.~\ref{fig_z2best}.

Across the entire frequency range and parameter space covered by our timing analysis, no other frequency had a power as high as $Z^2_\ast$. The peak was also the strongest for searches conducted with different photon patterns and frequency resolution. We found that for low frequency resolution searches, with steps of $8-10$\,$\mu$Hz, the exact frequency was missed and the power (although the highest in the search) is lower, $Z^2_\ast\sim33$. Best results were found when only single and double pn events were included in the search.

To investigate whether this signal is associated with unknown instrumental effects, we performed the same analysis on photons extracted from a few background regions, with roughly the same total number of counts as collected for \jten. We found that the maximum peaks (which occur at random frequencies) are always below the $1\sigma$ confidence level, considering the total frequency range of the analysis ($P=0.0114-1000$\,s and $\Delta\nu=87.72$\,Hz; see again Fig~\ref{fig_z2best}). The corresponding power of $Z^2_1$ statistics at $P=P_\ast$ is also very low in the background regions, $Z^2_{\rm \ast,bkg}\sim6$. Moreover, we looked for significant peaks around $P=P_\ast$ in most recent observations of other INSs performed in SW mode -- the 2009 observation of Calvera (\textsf{\small obsid}~0601180201) and the 2006 one of \rrat\ (\textsf{\small obsid}~0406450201). Peaks detected at our reported frequency are weak in both cases, $Z^2_{\rm \ast,Calvera}\sim0.7$ and $Z^2_{\rm \ast,RRAT}\sim5$, as shown in Fig.~\ref{fig_z2otherins}. 

Although the highest discovered in our search, the power (significance) of the $Z^2_1$ statistics at $P_\ast$ is sensitive both to the choice of energy band and the size of the extraction region (Fig.~\ref{fig_z2radband}).
In general, we found that potential pulsations from the source are likely to be smeared out by noise when the energy band is not restricted to below $\sim0.3$\,keV, where the background of the pn camera dominates. 
In particular, for searches with $\epsilon_l<0.25$\,keV, the peak at $P=P_\ast$ is below the $2\sigma$ confidence level (Fig.~\ref{fig_z2radband}; left). 
On the other hand, when low energy photons are discarded and the lower end of the energy band is not too restrictive (i.e. $\epsilon_l$ is between 0.3\,keV and 0.6\,keV), the signal is always detected above the $3\sigma$ confidence level. The loss of sensitivity detected when $\epsilon_l$ is above 0.6\,keV is due to the reduced number of counts (between 950 and 3750 counts in the $1-2.2$\,keV and $0.6-2.2$\,keV energy bands, respectively, compared to the more than 5000 counts in broader energy bands). 
The signal is not as sensitive to the upper end of the energy band, $\epsilon_u$, and is strongest at 2.2\,keV ($Z^2_\ast=49-50$ for $\epsilon_u$ between 1.9\,keV and 2.5\,keV; $\epsilon_l=0.36$\,keV). We note that by restricting ourselves to high-$S/N$ energy bands ($0.4-1.3$\,keV, $S/N>5$ and $0.33-1.5$\,keV, $S/N>3$), we obtain results that are no more reliable than more sensitive searches (i.e. with broader energy bands).

Our timing analysis carried out using photons extracted from circular regions with different radii revealed that the $Z^2_\ast$ power varies steeply between $17''$ and $20''$, reaching its maximum at $18.85''$ (Fig.~\ref{fig_z2radband}; right). For the $0.36-2.2$\,keV energy band, the signal is always above $3\sigma$ for radii between $16''$ and $20''$. Other energy bands also show an optimal extraction radius at the same value, $18.85''$.
\begin{figure}[t]
\begin{center}
\includegraphics[width=0.5\textwidth]{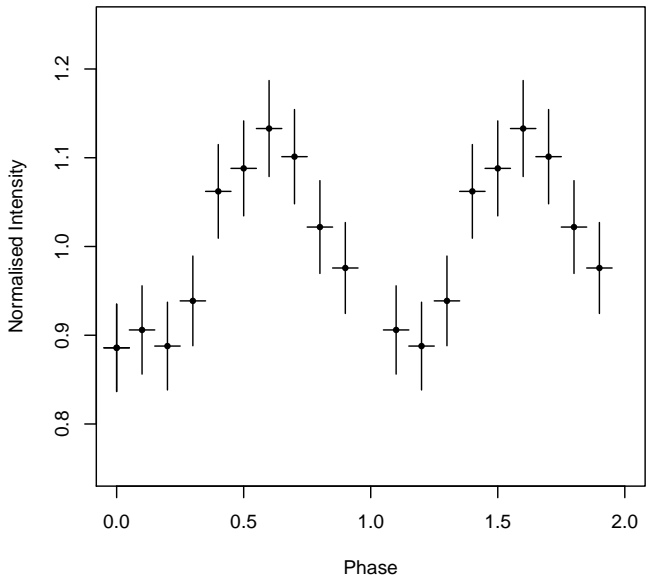}
\end{center}
\caption{Background-subtracted light curve for pn, folded at the spin period $P_\ast\sim18.6$\,ms. The energy band is $0.35-2$\,keV and the size of the extraction region is $18.85''$.}\label{fig_foldedP19}
\end{figure}

For the best choices of energy band and extraction region radius as discussed above, the significance of the pulsed signal is $4\sigma$, for $\Delta\nu=87.72$\,Hz and $\mathcal{N}=5.55\times10^6$ independent trials (the probability of the measured periodicity being spurious is $\sim6.334\times10^{-5}$). The corresponding pulsed fraction, defined as $p_{\rm f} = \big(2Z^2_{\rm max}/N_{\rm{ph}}\big)^{1/2}\times100\%$, where $N_{\rm{ph}}$ are the total number of counts in the search, is $\sim14\%$. A pn lightcurve, folded at $P_\ast$ and corrected for background counts and other effects, can be seen in Fig.~\ref{fig_foldedP19}. The pulse profile, as expected from the $Z^2_1$ search, is fairly sinusoidal.
\subsection{Spectral analysis\label{sec_spectra}}
Our analysis of archival data that serendipitously detected \jten\ \citepalias{pir09a} revealed a soft spectrum characterised by blackbody-like emission of an average temperature $kT=117\pm14$\,eV, hydrogen column density $\nh=(3.5\pm1.1)\times10^{21}$\,cm$^{-2}$, and stable observed flux of $f_\mathrm{X}=1.03(6)\times10^{-13}$\,erg\,s$^{-1}$\,cm$^{-2}$ (errors are at $3\sigma$). 

The analysis of AO9 data is based on source spectra extracted from a region of size $19''$, together with respective response matrices and ancillary files for each of the EPIC cameras (created using the \textsf{\small SAS} tools \textsf{\small rmfgen} and \textsf{\small arfgen}). Background spectra were extracted from regions as described in Sect.~\ref{sec_xmmobs_datareduction}. We limited the energy band of the pn camera to $0.3-2.2$\,keV, while the MOS cameras were analysed between 0.15\,keV and 2.2\,keV.

To increase the $S/N$ and use standard $\chi^2$-minimization techniques in \textsf{\small XSPEC~12.7.0}, spectra were rebinned using the \textsf{\small SAS} task \textsf{\small specgroup}. The spectral bins were grouped according to a minimum $S/N$ of three and we took care to oversample the instrument energy resolution at a given bin by less than a factor of three.
Our preliminary analysis had shown that the inclusion of data from the MOS2 camera considerably worsens the spectral fits and biases the parameter determination; we therefore only considered the pn and MOS1 cameras, which were fitted simultaneously to more tightly constrain the spectral parameters.

To account for interstellar absorption, we adopted the photoelectric absorption model and elemental abundances of \citet[][\textsf{\small tbabs} in \textsf{\small XSPEC}]{wil00}. We note that the source spectral parameters in \citetalias{pir09a} were derived by adopting a different abundance table \citep[that of][]{and89}, which produces a $\sim15\%$ higher value of the equivalent hydrogen column-density and a $\sim10\%$ softer blackbody temperature than those derived with \citeauthor{wil00} abundances (c.f. note~$^{\ref{foot_phabs}}$).

The results of our spectral fits are given in Table~\ref{tab_resultsspecfit}. We list for each fit (numbered [1--14] as reference for the text) the reduced chi-squared $\chi^2_\nu$, degrees of freedom and corresponding null-hypothesis probability, the equivalent hydrogen column density $\nh$, model-dependent parameters, and the unabsorbed source flux $F_{\rm X}$ in the $0.1-12$\,keV energy band. Unless otherwise noted, the errors in the parameters represent the $1\sigma$ confidence levels. 
\begin{table*}[t]
\caption{Results of spectral fits
\label{tab_resultsspecfit}}
\centering
\begin{tabular}{l c c l l l l l l}
\hline\hline
\multicolumn{1}{c}{Model} & $\chi^2_{\nu}$~(d.o.f.) & Null Hyp. & \multicolumn{1}{c}{$\nh$} & \multicolumn{3}{c}{Parameter} & \multicolumn{1}{c}{$F_{\rm X}$} & \multicolumn{1}{c}{Comment} \\
 & & (\%) & \multicolumn{1}{c}{(10$^{21}$\,cm$^{-2}$)} & \multicolumn{3}{c}{} & \multicolumn{1}{c}{(erg\,s$^{-1}$\,cm$^{-2}$)} & \\
\hline
\textsf{\scriptsize [1]~bbody} & 1.66(61) & 0.091 & $2.61_{-0.21}^{+0.22}$ & $kT_\infty$ & $138(3)$ & eV & $6.8_{-0.7}^{+0.9}\times 10^{-13}$ & {\tiny \citeauthor{wil00} abundances} \\[0.1cm]
\textsf{\scriptsize [2]~bbody} & 2.22(61) & $1.3\times10^{-5}$ & $2.76(23)$ & $kT_\infty$ & $152(3)$ & eV & $5.9(4)\times10^{-13}$ & {\tiny \citeauthor{ham07} abundances} \\[0.1cm]
\textsf{\scriptsize [3]~bbody} & 1.21(60) & 13 & $3.31^{+0.28}_{-0.27}$ & $kT_\infty$ & $124(4)$ & eV & $1.20(22)\times10^{-12}$& {\tiny $Z_{\rm O}=1.66(12)$ (solar)} \\[0.1cm]
\hline
\textsf{\scriptsize [4]~nsa $B_0$}    & 1.64(61) & 0.114 & $4.25_{-0.16}^{+0.21}$ & $T_{\rm eff}$ & $4.59_{-0.15}^{+0.11}\times10^5$ & K & $3.3_{-1.0}^{+1.5}\times10^{-12}$ & {\tiny $M^{\star}=1.4$\,M$_\odot$, $R^{\star}=10$\,km} \\[0.1cm]
\textsf{\scriptsize [5]~nsa $B_0$}    & 1.70(59) & 0.067 & $4.25(18)$ & $T_{\rm eff}$ & $4.5_{-0.4}^{+0.3}\times10^5$ & K & $3.3_{-1.4}^{+1.7}\times10^{-12}$ & {\tiny $M=1.3^{+0.3}_{-0.8}$\,M$_\odot$, $R\sim10$\,km\,\tablefootmark{a}} \\[0.1cm]
\textsf{\scriptsize [6]~nsa $B_{12}$} & 1.80(61) & 0.013 & $4.20^{+0.22}_{-0.4}$ & $T_{\rm eff}$ & $7.1(3)\times10^5$ & K & $2.8_{-0.9}^{+1.4}\times10^{-12}$ & {\tiny $M^{\star}=1.4$\,M$_\odot$, $R^{\star}=10$\,km} \\[0.1cm]
\textsf{\scriptsize [7]~nsa $B_{12}$} & 1.43(59) & 1.673 & $2.15_{-0.20}^{+0.27}$ & $T_{\rm eff}$ & $3.0_{-0.5}^{+0.7}\times10^6$ & K & $4_{-3}^{+5}\times10^{-13}$ & {\tiny $M\sim2$\,M$_\odot$, $R\sim7$\,km\,\tablefootmark{b}} \\[0.1cm]
\textsf{\scriptsize [8]~nsa $B_{13}$} & 1.98(61) & $7.6\times10^{-4}$ & $4.66_{-0.4}^{+0.21}$  & $T_{\rm eff}$ & $6.81_{-0.3}^{+0.27}\times10^5$ & K & $3.9_{-1.6}^{+3}\times10^{-12}$ & {\tiny $M^{\star}=1.4$\,M$_\odot$, $R^{\star}=10$\,km} \\[0.1cm]
\hline
\textsf{\scriptsize [9]~powerlaw} & 2.37(61) & $9.1\times10^{-7}$ & $9.9(4)$ & $\Gamma$ & $8.58(26)$ & & $8(5)\times10^{-7}$ & {\tiny \citeauthor{wil00} abundances}\\[0.1cm]
\hline
\textsf{\scriptsize [10]~bbody-gauss}  & 1.05(57) & 38 & $2.7(3)$     & $kT_\infty$ & $135_{-6}^{+8}$ & eV & $7.5_{-1.6}^{+2.1}\times10^{-13}$ & {\tiny $Z_{\rm O}=1.64(15)$ (solar)} \\
                                       &          &    &              & $\epsilon$  & $1.36_{-0.05}^{+0.03}$ & keV & & {\tiny $\sigma=0.14 (1\sigma), \textrm{EW}=91$\,eV} \\[0.1cm]
\textsf{\scriptsize [11]~bbody-2*gauss}& 1.01(58) & 46 & $2.61^\star$ & $kT_\infty$   & $134.9(2.4)$ & eV & $7.3(4)\times10^{-13}$ & {\tiny \citeauthor{wil00} abundances} \\
                                       &          &    &              & $\epsilon_1$  & $0.611_{-0.018}^{+0.020}$ & keV & & {\tiny $\sigma=0.1^\star,\textrm{EW}=71$\,eV} \\
                                       &          &    &              & $\epsilon_2$  & $1.35(4)$ & keV & & {\tiny $\sigma=0.1^\star,\textrm{EW}=86$\,eV} \\[0.1cm]
\textsf{\scriptsize [12]~bbody-2*gauss}& 1.00(57) & 48 & $2.2(3)$     & $kT_\infty$   & $141(6)$ & eV & $5.7_{-0.9}^{+1.2}\times10^{-13}$ & {\tiny \citeauthor{wil00} abundances} \\
                                       &          &    &              & $\epsilon_1$  & $0.597_{-0.022}^{+0.023}$ & keV & & {\tiny $\sigma=0.1^\star,\textrm{EW}=71$\,eV} \\
                                       &          &    &              & $\epsilon_2$  & $1.35(3)$ & keV & & {\tiny $\sigma=0.1^\star,\textrm{EW}=104$\,eV} \\[0.1cm]
\textsf{\scriptsize [13]~bbody+bbody}  & 1.17(59) & 17 & $5.2(6)$     & $kT_1^\infty$ & $37(4)$  & eV & $1.3_{-0.8}^{+27}\times10^{-10}$ & {\tiny $R_1^\infty\sim640$\,km\tablefootmark{c}} \\
                                       &          &    &              & $kT_2^\infty$ & $119(4)$ & eV & & {\tiny $R_2^\infty\sim6$\,km\tablefootmark{c}} \\[0.1cm]
\textsf{\scriptsize [14]~bbody+pow}    & 1.66(61) & 0.091 & $2.61^\star$ & $kT_\infty$ & $137.9^{+1.2}_{-1.4}$ & eV & $6.8(3)\times10^{-13}$ & {\tiny \citeauthor{wil00} abundances} \\
                                       &          &       &              & $\Gamma$    & $1.7^\star-2.1^\star$ &    &  & {\tiny $F_{\rm X}^{\rm pl}=(6.8-7.3)\times10^{-15}(3\sigma)$\tablefootmark{d}} \\[0.1cm]
\hline
\end{tabular}
\tablefoot{Parameters marked with a star are held fixed during fitting.\\
\tablefoottext{a}{Stellar radius unconstrained by the model.}
\tablefoottext{b}{Stellar mass and radius unconstrained by the model.}
\tablefoottext{c}{Computed for distance $d=2.3$\,kpc.}
\tablefoottext{d}{Upper limit to the unabsorbed flux of the power-law component, in units of erg\,s$^{-1}$\,cm$^{-2}$ ($0.1-12$\,keV).}
}
\end{table*}

We tested several models and found that the emission of \jten\ is consistent with previous analysis based on archival data\footnote{The best-fit blackbody model gives $kT_\infty=122(4)$\,eV and $\nh=3.20^{+0.27}_{-0.26}\times10^{21}$\,cm$^{-2}$, with the abundance table of \citet{and89}.\label{foot_phabs}}. However, the much higher quality of our statistics revealed that a single component is hardly satisfactory in describing the source energy distribution. The best-fit single-component model consists of either a blackbody or a non-magnetized neutron star atmosphere. 
The null-hypothesis probability, which gives the likelihood of having a larger value of $\chi^2_\nu$ by chance, assuming that the model is correct, is low -- at best lower than $2\%$ for all single-component models ([1--9] in Table~\ref{tab_resultsspecfit}). The best-fit power-law model [9] is significantly worse than the thermal models, with a null-hypothesis probability $\ll1\%$. The photon index $\Gamma\sim9$ is unreasonably steep and the model is four times more absorbed than the thermal ones. Other models -- bremsstrahlung and optically thin thermal plasma \citep{ray77} -- show even poorer fits. We discuss below the results of the thermal models in more detail.

The best-fit absorbed blackbody [1] has $kT_\infty=138(3)$\,eV and $\nh=2.61_{-0.21}^{+0.22}\times10^{21}$\,cm$^{-2}$, with $\chi^2_\nu\sim1.7$ for 61 degrees of freedom. The radiation radius, as seen by an observer at infinity, is $R_\infty \sim (3-5) (d/d_{\rm Car})^{-1}$\,km, where the distance to \jten\ is normalised to that of \etacar, $d_{\rm Car}=2.3$\,kpc \citep{smi06a}. In this estimate, we included the flux uncertainty from the abundance tables. Although this radius is smaller than that of a canonical neutron star, it agrees with those measured for the \msev: their redshifted radiation radii, as derived from X-ray blackbody fits and distance estimates, are in the range of, roughly, 2\,km to 7\,km (using the distance estimates and source parameters from e.g. \citealt{kap08a}; see also Sect.~\ref{sec_discLinks}).

A neutron star atmosphere model \citep[][\textsf{\small nsa}]{zav96,pav95} describes the data statistically nearly as well [4--8]. A stellar mass of $M_{\rm ns}=1.4$\,M$_\odot$ and radius of $R_{\rm em}=10$\,km were first assumed [4,6,8], then allowed to vary to check for an improved fit [5,7]. The magnetic field strength was held fixed during fitting, and we tested the values $B=0$\,G [4--5], $B=10^{12}$\,G [6--7], and $B=10^{13}$\,G [8]. 
The best-fit models typically fit a 60\% higher $\nh$ than the blackbody and a much softer temperature, $kT_\infty\sim30$\,eV. For the typical absorption in the direction of the Carina Nebula, the implied luminosity distance is implausibly small, $d\sim220$\,pc.
Poorer fits were found when the magnetic field was included and with increasing field intensity. Although the fit was generally not improved when the neutron star mass and radius were allowed to vary, we found a statistically tighter fit for a magnetic field strength of $B=10^{12}$\,G and with freely varying $M_{\rm ns}\sim2$\,M$_\odot$ and $R_{\rm ns}\sim7$\,km (null-hypothesis probability of $1.7\%$; [7]). In this case, $\nh$ and $kT_\infty$ were more similar to the most closely fitting blackbody values. However, the stellar mass and radius were largely unconstrained by the model, and the implied source distance is much larger than expected from the measured $\nh$ ($d\sim7.5$\,kpc). 

\begin{figure*}[t]
\begin{center}
\includegraphics*[width=0.49\textwidth]{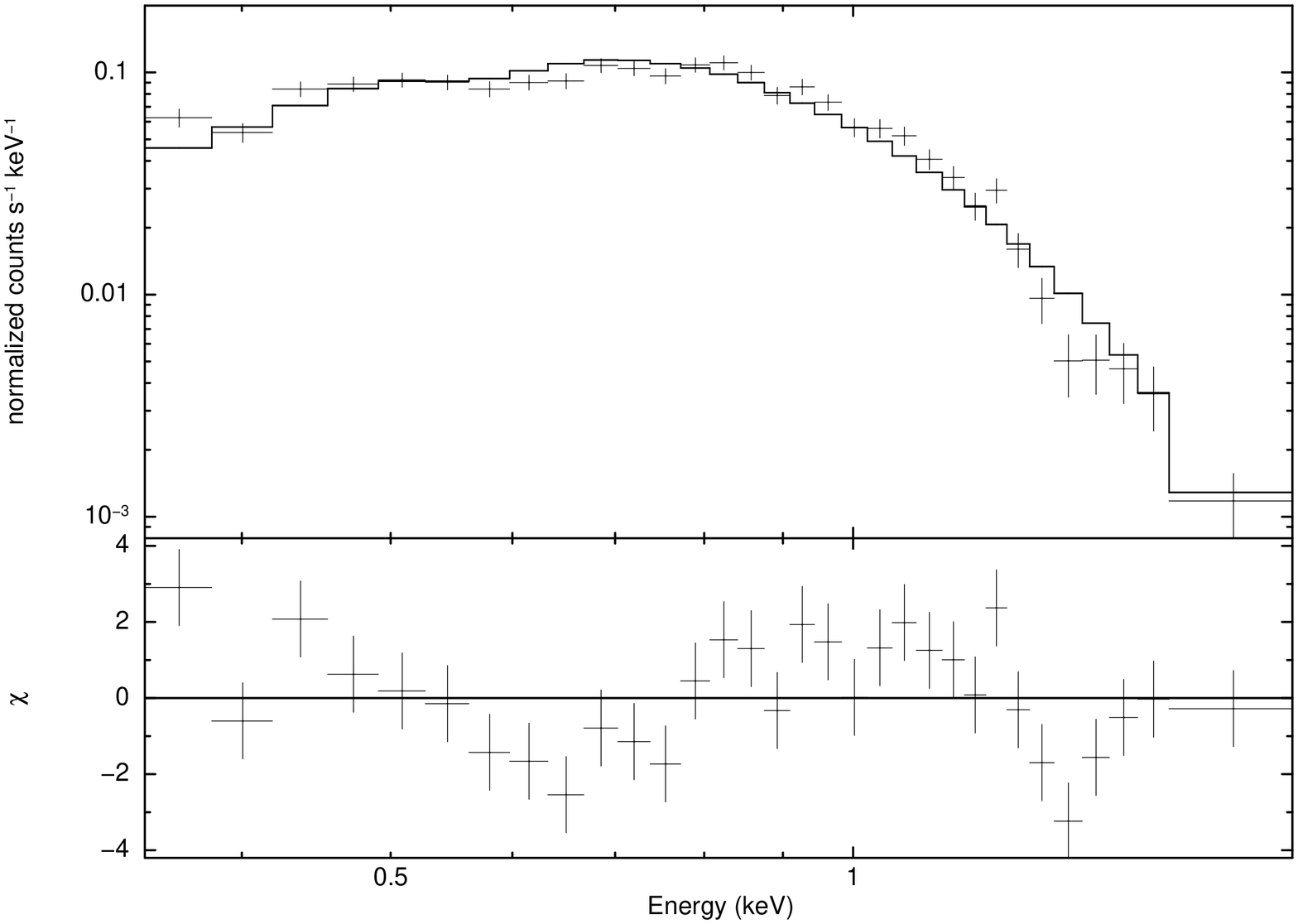}\hspace{5pt}
\includegraphics*[width=0.49\textwidth]{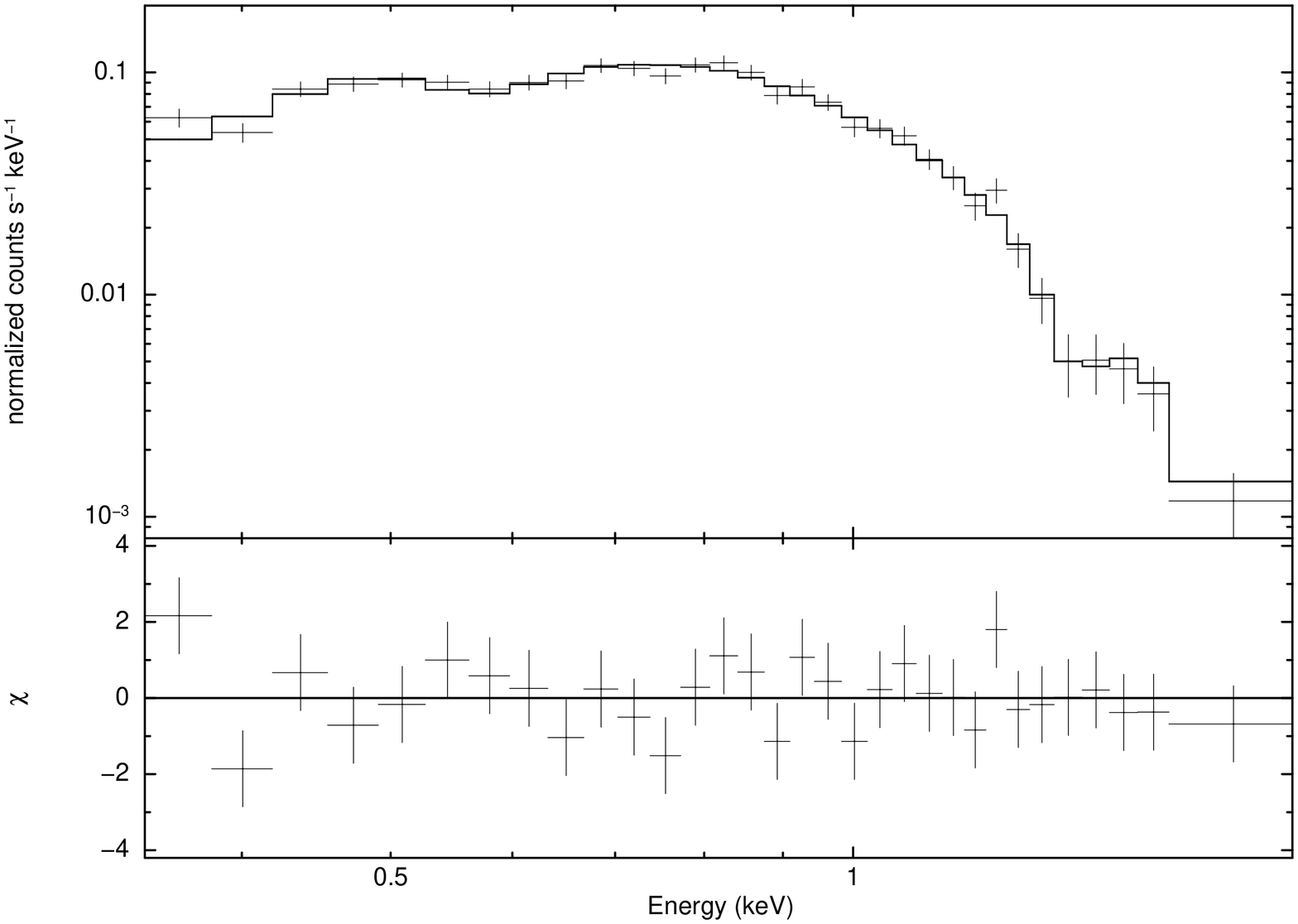}
\end{center}
\caption{Results of spectral fitting of source \jten. We show the pn data and folded best-fit blackbody model, with residuals. \textit{Left.} Simple blackbody. \textit{Right.} Best-fit with oxygen overabundance and a Gaussian absorption line at energy $\epsilon\sim1.35$\,keV.}\label{fig_spec}
\end{figure*}

The best-fit blackbody/neutron star models invariably show residuals around $0.6-0.7$\,keV and $1.3-1.4$\,keV (see Fig.~\ref{fig_spec}; left). We note that these features are independent of the choice of source and background regions used to create the spectra (thus exclude improper background subtraction), of the EPIC instrument (they are seen in the data from the three EPIC cameras, but are more pronounced in pn), as well as of the choice of cross-sections and abundance tables available in \textsf{\small XSPEC}. Before adding yet more complexity to the spectral model, we investigated whether the residuals could be due to a local elemental overabundance in the Carina Nebula. 

Recent studies \citep[e.g.][and references therein]{tow11b} have investigated the properties (morphology, emissivity, and elemental abundances) of the hot plasma known to permeate the Carina Nebula. The interaction of the plasma with cold molecular gas is likely responsible for the diffuse soft X-ray emission first detected with the Einstein Observatory (\citealt{sew79}; see Sect.~\ref{sec_discCarina}). Its composition and, in particular, enhanced abundances of key elements indicate whether this soft emission originates in either the wind of the current population of more than 70 massive stars \citep{smi06b} and/or in one or more supernova explosions \citep{ham07}. 

We tested the hypothesis of a local elemental overabundance using the variable photoelectric absorption model \textsf{\small vphabs} convolved with a blackbody in \textsf{\small XSPEC}. We first attempted to fit the spectra by fixing the elemental abundances at the best-fit values found by \citet{ham07}, with freely varying $\nh$ and $kT$ [2]. However, the result was considerably worse than that obtained with solar abundances from \citeauthor{wil00} Moreover, the residuals remained roughly at the same energies. We then tested for the overabundance of all elements with significant transitions in the range of 0.2\,keV to 2\,keV (C, N, O, Ne, Mg, \ldots). We restricted our test to abundance values in the range $1<Z<5$ in solar units to avoid both subsolar and arbitrarily high metal abundances \citep[see e.g.][for a discussion]{tow11b}. 

Interestingly, we found that, whereas most of the elemental abundances (e.g. C, N, S) remained insensitive to the fit and others (most notably, Mg) were pegged to the hard end of the allowed interval in abundance, the only element that was tightly constrained was oxygen, with $Z_{\rm O}=1.66(12)$ in solar units [3]. By allowing $Z_{\rm O}$ to be extrasolar, we obtained a much closer agreement between data and model for energies below 1\,keV. However, an inspection of the residuals revealed that the feature at energy $\sim1.3$\,keV remained in spite of the improved fit and could only be accounted for with a Gaussian line in absorption. We indeed verified that elements with lines around this energy tended to be arbitrarily overabundant in the model. 

The inclusion of a Gaussian line in absorption improves the fit considerably [10]. Allowing all parameters to vary freely, we obtained a good fit with $\chi^2_\nu\sim1$ and a much higher null-hypothesis probability. The best-fit model and its residuals can be seen in the right panel of Fig.~\ref{fig_spec}. This model -- with $\nh$ and $kT$ consistent with the (simple) blackbody values, an oxygen overabundance of $Z_{\rm O}=1.64(15)$, and a line in absorption at energy $\epsilon=1.36^{+0.03}_{-0.05}$\,keV ($\sigma=0.14$\,keV and equivalent width of $\textrm{EW}=91$\,eV) -- reproduces the observed spectral energy distribution of \jten\ very well. We verified that the results ($\epsilon, Z_{\rm O}$) are insensitive to any changes in $\nh$. 

For completeness, we tested other multicomponent models, namely a double blackbody [13], a blackbody with multiple Gaussian absorption lines (with fixed solar abundances, [11--12]), and a blackbody with a power-law tail [14]. The fit results are also listed in Table~\ref{tab_resultsspecfit}.
All parameters in the double blackbody [13] were allowed to vary during fitting. The softer blackbody temperature $kT_1$ was restricted to between 5\,eV and 120\,eV and the hotter one $kT_2$ to between 0.1\,keV and 1\,keV. The best fit ($\chi^2_\nu\sim1.2$) gives temperatures of 37\,eV and 119\,eV. Relative to the single-blackbody fit, the addition of the extra component provides a closer fit to the softer part of the spectrum, eliminating the ``excess'' detected previously and lowering the value of $\chi^2$. However, the two features at energies $\sim0.6$\,keV and $\sim1.3$\,keV discussed before remain. Given the substantial interstellar medium absorption of the model ($\nh\sim5\times10^{21}$\,cm$^{-2}$), the normalisation of the soft component is forcibly very high, two orders of magnitude higher than the hot component. The redshifted emission radius, at a distance of 2.3\,kpc, is $R_1^\infty\sim640$\,km and $R_2^\infty=6$\,km. This argues against the physical viability of the model.

To test for the presence of multiple absorption lines, we adopted the following strategy: first, we fitted the data holding the column density fixed at the best single-blackbody value, $\nh=2.61\times10^{21}$\,cm$^{-2}$, also keeping the Gaussian $\sigma$ parameter fixed at a value of 0.1\,keV [11]. We then allowed to vary either $\sigma$ or $\nh$. Finally, we fitted the data allowing all parameters (but the equivalent widths) to vary freely [12]. For all fits, the blackbody temperature was restricted to within 10\,eV and 300\,eV and the line energy to within 0.1\,keV and 2\,keV.
Spectral fits show good results with two absorption lines invariably at energies around $0.58-0.61$\,keV and $1.35-1.38$\,keV. The presence of the lines does not lead to any significant changes in the best-fit values of $kT$ and $\nh$. When $\nh$ varies freely, the blackbody is hotter relative to the single-component temperature ($kT\sim140$\,eV) and the column density is lower, at $\nh\sim2.2\times10^{21}$\,cm$^{-2}$ [12]. 

We note that \citet{ham09} reported evidence of absorption features in the combined spectra of \jten, namely a slight dip around 0.9\,keV and a strong one at 0.6\,keV. As noted by the authors, the latter is likely to be produced by edge absorption of oxygen in the interstellar medium or in the detector response. We found no evidence in AO9 data to confirm the dip at 0.9\,keV, which might have been produced during the combination of both the spectra and calibration files of the three EPIC detectors taken at various off-axis angles and with different instrument configurations.

The blackbody model in combination with a power law [14] was tested to derive the upper limits to the detection of a non-thermal component extending towards higher energies, as usually seen in the emission of middle-aged pulsars dominated by soft thermal components (age $\sim$ few $10^5$\,yr, e.g. the ``Three Musketeers''); in these objects, power-law tails with $\Gamma=1.7-2.1$ were detected at very low flux levels ($0.3-1.7\%$ of the source luminosity; \citealt{luc05}).
To derive upper limits to the detection of a hard non-thermal component, we fitted a power law with photon indices $\Gamma=1.7$ and $\Gamma=2.1$, in addition to the dominant blackbody component. The photon index and a column density of $\nh=2.6\times10^{21}$\,cm$^{-2}$ were kept fixed during the fitting. We found that the best-fit blackbody parameters were unaffected by the inclusion of the power-law component. Relative to the single-blackbody model, the fit did not improve and the non-thermal component was found to contribute at most $1-1.1\%$ ($3\sigma$ confidence level, $0.1-12$\,keV range) of the unabsorbed flux of the source.
\section{ESO-VLT observations\label{sec_opt}}
\begin{table}[t]
\caption{Description of the optical \eso\ observations of \jten
\label{tab_logopt}}
\centering
\begin{tabular}{c c c c c c}
\hline\hline
\multicolumn{1}{c}{Night} & Period & Exposures & $t_{\rm exp}$\tablefootmark{a} & FWHM & Airmass \\
 & & & (s) & (arcsec) & \\
\hline
2009-02-21 & P82 & $3\times\textrm{V}$         & 3300 & 0.50 & 1.22\\
2010-04-07 & P85 & $6\times\textrm{H$\alpha$}$ & 6264 & 1.16 & 1.23\\
2010-06-06 & P85 & $3\times\textrm{H$\alpha$}$ & 3132 & 0.67 & 1.26\\
\hline
\end{tabular}
\tablefoot{The exposure times, seeing, and airmasses are averages per filter.
\tablefoottext{a}{Total exposure time per filter.}
}
\end{table}
Optical follow-up observations of \jten\ with \eso\ and \soar\ in 2007 and 2008 revealed no counterparts brighter than $m_{\rm R}=25$, $m_{\rm B}=26$, and $m_{\rm V}=25.5$ \citepalias[$2.5\sigma$;][]{pir09a}. To derive a deeper limit to the brightness of the optical counterpart, we obtained further \eso\ V-band exposures in 2009. The V band is the optimal choice for deep imaging because it excludes some of the strongest lines found in \ion{H}{ii} regions (Balmer emission as well as [\ion{N}{ii}] and [\ion{S}{ii}] emission lines), thus minimizing contamination from bright nebular lines. 
Additionally, we obtained deep H$\alpha$ imaging with \eso\ in 2010, in order to investigate a possible Balmer-dominated nebula near the position of the X-ray source. 
\begin{figure*}[t]
\begin{center}
\includegraphics[width=\textwidth]{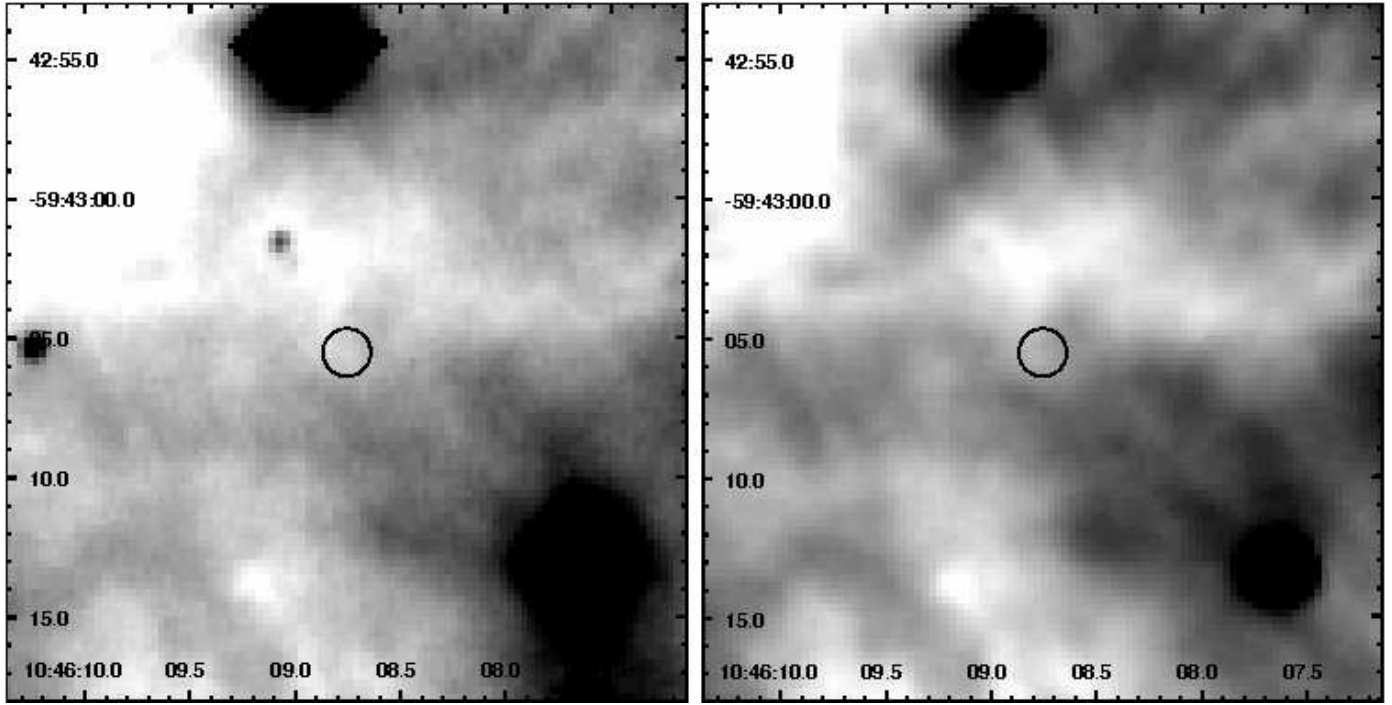}
\end{center}
\caption{V (left) and H$\alpha$ (right) images of the field of source \jten. An inverted colour map is used, i.e. brighter objects are darker. The 90\% confidence level error circle on the position of the source is of size $0.86''$. The images were smoothed using a Gaussian filter of one pixel in size.}\label{fig_opt1046}
\end{figure*}
\subsection{Observations and data reduction}
The VLT 2009 and 2010 observations were obtained under photometric and dark sky conditions for a total $\sim3.53$\,h of observing time (see the log of observations in Table~\ref{tab_logopt}). The totality of the proposed observing time was executed in two different observing periods (P82 and P85). On both occasions, one of the FOcal Reducer low/dispersion Spectrographs \citep[FORS;][]{app98} was used. 

The FORS detector consists of a mosaic of two $2\textrm{k}\times4\textrm{k}$ CCDs of pixel size 15\,$\mu$m. In the 2010 observations (P85), we used FORS2, the camera mounted on UT1, which is optimized for the red band (MIT CCDs), while the 2009 observations of P82 were carried out with FORS1 (E2V CCDs) at UT2. The spectrograph FORS1 is most sensitive to the blue range. 
The detector provides imaging on a pixel scale of $0.25''$\,pixel$^{-1}$ (field-of-view of $6.8'\times6.8'$) using standard-resolution collimator and default readout mode ($2\times2$ binning). 
Dithering patterns with offsets of $5''$ were chosen for the observations (with $3\times\textrm{V}$ and $9\times\textrm{H$\alpha$}$ exposures; see Table~\ref{tab_logopt}). The seeing was excellent during the 2009 V observations; unfortunately, it was above the constraint of $0.8''$ FWHM during the observing night of April 2010, when most (67\%) of our P85 observations were executed. 

Data were reduced in two steps. Firstly, we applied \textsf{\small IRAF~v2.14} \citep{tod86} to process raw frames; secondly, \textsf{\small EsoRex~4.3.5}\footnote{http://www.eso.org/sci/software/cpl/esorex.html} and \eso\ recipes were used in order to recreate the instrument pipeline. For both sets of software, standard procedures were adopted and carried out by treating each FORS CCD chip independently. 
The flat-fielded scientific exposures were combined in order to remove both cosmic rays and bad/hot pixels and increase the $S/N$ of the data. We verified that both (\textsf{\small IRAF}/\textsf{\small EsoRex}) reductions yielded very similar results in the analysis, which was also true when compared to the results obtained with the pipeline-reduced scientific frames distributed by \eso. 
\subsection{Data analysis\label{sec_optanalysis}}
The transformation of the $m_{\rm V}$ instrumental magnitudes to the standard photometric system was done using the zero-point, extinction coefficient and colour term obtained on the night of the observation, as provided by the \eso\ calibration webpages\footnote{http://www.eso.org/observing/dfo/quality}. The H$\alpha$ images were flux-calibrated using observations of a standard star observed in the same night of the observation (LTT\,4816; \citealt{ham94}).
The astrometric calibration of the stacked science frames was performed using the USNO-B1.0, 2MASS, and GSC2 catalogues and the \textsf{\small GAIA~4.2-1} software\footnote{http://star-www.dur.ac.uk/$\sim$pdraper/gaia/gaia.html}. 
In general, our astrometric errors are of $\sim$\,$0.15''$ or better.
We adopted standard PSF fitting, as implemented in the \textsf{\small daophot} \citep{ste87} package for the \textsf{\small IRAF} environment, to measure magnitudes. 

The deep V observation failed to detect the optical counterpart to \jten: as can be seen in the optical image of Fig.~\ref{fig_opt1046}, no object lies within $\sim$\,$4.3''$ ($\gtrsim$\,$5\sigma$) of the source position. 
To test the detection limit specifically in the region covered by the error circle of the X-ray source, we added images of synthetic stars of progressively fainter magnitudes to a small section of the combined V image and analysed the combined images using PSF fitting \citep[see][for details]{pir09b}. We defined the limiting magnitude of the 2009 observations as $m_{\rm V}=27.6(5)$ ($2\sigma$), which corresponds to the detected magnitude of the faintest synthetic star successfully measured in the image. 
Taking the limiting magnitude as an upper limit to the brightness of the optical counterpart of \jten, the implied X-ray-to-optical flux ratio is $\log(F_{\rm X}/F_{\rm V})>3.8$, corrected for photoelectric absorption and interstellar extinction. The de-reddened V flux is derived using the X-ray absorption (Sect.~\ref{sec_spectra}) and the optical extinction, $A_{\rm V} \sim 1.51$, computed with the \citet{pre95} relation.
\subsection{Searching for a bow-shock or photoionisation nebula}
\begin{figure}[t]
\begin{center}
\includegraphics[width=0.49\textwidth]{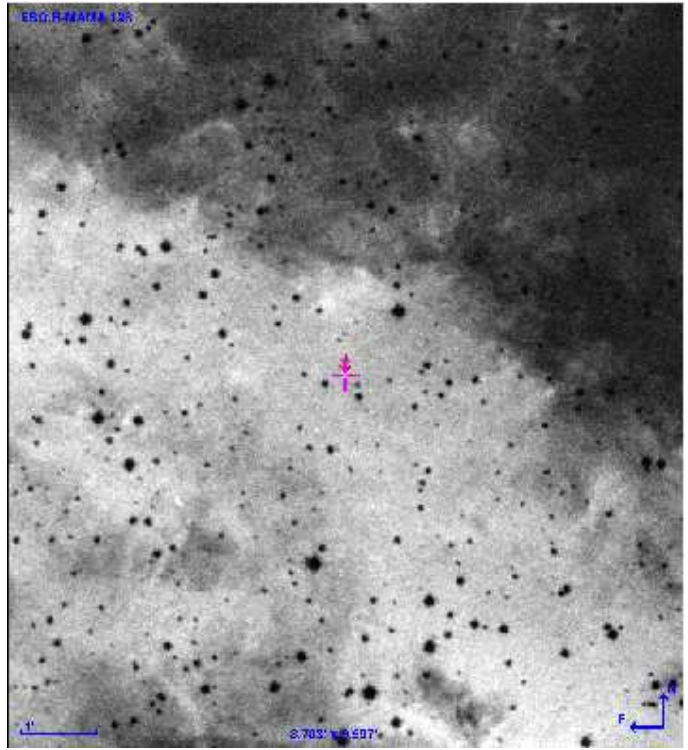}
\end{center}
\caption{ESO-MAMA wide-field R band image of the area around \jten\ (marked by a cross), in the direction of the Carina Nebula. The young OB associations Tr\,16 and Tr\,14, as well as \etacar, are located NW of the source, towards the bright diffuse emission.}\label{fig_fieldRMAMA}
\end{figure}

The dense environment of the Carina Nebula and the possibly lowly-ionised local medium surrounding \jten\ favour the formation of a bow-shock or of a photoionised nebula that could be detected in the optical. The fraction of neutral hydrogen in the nebula depends on the distance from the sources of ionising radiation, mostly OB stars present in the young stellar clusters Tr\,16 (where \etacar\ is) and Tr\,14, and on the clumpiness of the local medium. \jten\ is located in a region of low diffuse background relative to other parts of the nebula, and is apparently behind an ionisation front from Tr\,16 (see Fig.~\ref{fig_fieldRMAMA}). Therefore, most of the local interstellar medium may still be neutral, which favours the detection of an H$\alpha$ signature around the position of the presumed rapidly-moving neutron star. 

A Balmer-dominated nebula was seen around a number of radio pulsars \citep[e.g.][]{gae06} and one (possibly two, see \citealt{mot05} and the case of \magonesix) of the \msev, \magoneeig\ \citep{ker01}.  Two possible scenarios are invoked to explain the presence of such a nebula: bow-shocks \citep[e.g.][]{cha02} and photoionisation by X-rays \citep{bla95}. The shape and emissivity of the nebula can provide invaluable information about the INS velocity, direction of motion, and rotational energy losses, allowing one to constrain its age and evolutionary state. 

Motivated by the above, we estimated the expected H$\alpha$ flux from \jten\ for both scenarios, considering angular (projected) separations of the apex of the nebula relatively to the neutron star position (stand-off radius $R_0$) in the range from $0.5''$ to $5''$. Simulations using the photoionisation code \textsf{\small XSTAR}\footnote{http://heasarc.gsfc.nasa.gov/docs/software/xstar/xstar.html} predicted an H$\alpha$ flux (taking into account the interstellar extinction) of $f_{{\rm H}\alpha}\sim10^{-17}-10^{-16}$\,erg\,s$^{-1}$\,cm$^{-2}$\,arcsec$^{-2}$, for a local density of $50$\,cm$^{-3}$ \citepalias[e.g.][]{pir09a} assuming the spectral properties of \jten\ (Sect.~\ref{sec_spectra}). The resulting flux range is consistent with the expectations of \citet{bla95} for a similar neutron star luminosity and medium density, and with that measured for the \msev\ \magoneeig. 

Alternatively, in the bow-shock scenario, a neutron star of velocity $v_{\rm ns}=10-100$\,km\,s$^{-1}$ moving through a uniform medium of density $n=50$\,cm$^{-3}$, where (conservatively) half of the hydrogen atoms is still neutral, requires a spin-down luminosity of $\dot{E}>10^{34}$\,erg\,s$^{-1}$ to create bow-shocks with stand-off radii in the range $R_0=0.5''-5''$ \citep[see the formulation of e.g.][]{cor93a}. These shocks correspond to an H$\alpha$ flux of $f_{{\rm H}\alpha}\sim10^{-15}-10^{-14}$\,erg\,s$^{-1}$\,cm$^{-2}$\,arcsec$^{-2}$, accounting for absorption. If, conversely, the spin-down luminosity of \jten\ is lower, more typical of those of the \msev\ ($\dot{E}\sim10^{31}$\,erg\,s$^{-1}$), the H$\alpha$ nebula is then expected to be less bright, $f_{{\rm H}\alpha}\sim4\times10^{-17}$\,erg\,s$^{-1}$\,cm$^{-2}$\,arcsec$^{-2}$, and (at $d=2.3$\,kpc) nearly angularly coincident with the position of the neutron star.
In any case, these figures indicate that the flux in the Balmer line may be high enough to allow detection with $\sim3$\,h of VLT observations, for both (photoionisation/bow-shock) scenarios.

Previous H$\alpha$ imaging of the field of the X-ray source, obtained with \soar\ in 2008, was badly affected by fringing \citepalias{pir09a}. Our new VLT observations do reveal some extended structure in the vicinity (within $1"$) of the position of the X-ray source (Fig.~\ref{fig_opt1046}; right). The H$\alpha$ flux is $f_{{\rm H}\alpha}\sim3.6\times10^{-15}$\,erg\,s$^{-1}$\,cm$^{-2}$\,arcsec$^{-2}$ -- at least one order of magnitude brighter than that expected if the nebular emission was created by X-ray photoionisation as previously discussed, but still formally consistent with the expectations of the bow-shock scenario for $\dot{E}>10^{34}$\,erg\,s$^{-1}$ (typical of young rotation-powered pulsars). However, given the presence of similar diffuse emission in other parts of the nebula, it is unlikely that the observed structure is indeed related to the neutron star.
\section{Gamma-ray analysis\label{sec_gamma}}
Our measurement of the neutron star spin period permits us to search for pulsed emission also in $\gamma$-rays, using data from the \agile\ and \fermi\ missions. This is particularly interesting given the large time spans of data that can be used to constrain the period derivative of the pulsar. Although no plausible $\gamma$-ray counterpart is present within $>5\sigma$ of the position of the X-ray source in either the \agile\ Bright Source Catalogue \citep{pit09} or the Second \fermi\ Catalogue \citep[2FGL;][]{2fgl11}, a search for a periodic signal can be more sensitive than spatial detections \citep[see][and references therein, for a discussion]{zan11}.

Data reduction was performed by making use of the software specific to each telescope and the whole available data span of both satellites. Public data from the \agile\ mission were accessed via the \textit{Agenzia Spazionale Italiana} Science Data Center (ASDC) Multi-Mission Interactive Archive\footnote{ http://www.asdc.asi.it/mmia/index.php?mission=agilemmia}. The analysis covers the time span from 2007 July 13 (MJD~54294) to 2011 April 5 (MJD~55656), which amounts to $169,289$ photons extracted around $5^\circ$ (in order to match the point spread function of the instruments at energies $E>100$\,MeV) from the position of \jten, as reported in Table~\ref{tab_coordinates}. We applied the \agile\ pulsar pipeline to perform data selection and barycentering. For details of the \agile\ observing strategy, timing calibration, and $\gamma$-ray pulsar analysis, we refer the reader to \citet{pel09b,pel09a}. 

The public archive of the Fermi Science Support Center\footnote{http://fermi.gsfc.nasa.gov/cgi-bin/ssc/LAT/LATDataQuery.cgi} was accessed to retrieve \fermi\ data. Again defining a region-of-interest of $5^\circ$ around the position of the source, $760,327$ photons with energy higher than 100\,MeV and registered between 2008 August 5 (MJD~54683) and 2011 March 30 (MJD~55650) were extracted in the timing analysis, using the \fermi\ Science Tool\footnote{Available at http://fermi.gsfc.nasa.gov/ssc/data/analysis/software} \textsf{\small gtselect}. Good-time intervals and data filters recommended by the \fermi\ Team were also applied with \textsf{\small gtmktime}. Finally, the photons were barycentred using the \textsf{\small gtbary} tool. 

We searched for periodicities using as a starting point the spin period of the source observed in X-rays (Sect.~\ref{sec_timing}). The two data sets (\agile/\fermi) were used separately and simultaneously. A pulsed signature was searched in frequency domain close to the nominal frequency $\nu_\ast=53.645804$\,Hz and around it, using a frequency step of the order $1/T_{\rm s}$ (where $T_{\rm s}$ is the time span of the analysed data set). To look for significant pulsations, we applied a Pearson $\chi^2$ test to the ten-bin pulsed profile resulting from each search trial. We also applied a (bin-independent) $Z^2_n$ test \citep{buc83} to each analysed data set. 

It is worth noting that the restricted information available for \jten\ greatly enhances the search in terms of parameter space.
The number of trials, given the long $T_{\rm s}$ for both the \agile\ and \fermi\ data sets, significantly increases, even when searching only around $1\sigma$ in frequency from the nominal value. As a first approach, the first and second period derivatives were set to zero and not allowed to vary. A wider search was then performed across a small range of period derivatives. 
Furthermore, to extend the search to a broader frequency range (around $10\sigma$ of $\nu_\ast$), the time span of the analysis was restricted to cover only the photons registered in December 2010, when the X-ray observations took place.

No significant periodicity ($>4\sigma$) was found. Although tentative pulsations result from individual searches, overall inconsistencies prevent a secure detection from being claimed in $\gamma$-rays.
Hence, we set a $4\sigma$ upper limit of $F_\gamma<4\times10^{-8}$\,ph\,s$^{-1}$\,cm$^{-2}$ at energies above 100\,MeV, for a pulsed $\gamma$-ray signal from \jten.
\section{Discussion\label{sec_disc}}
We first summarize our results and then discuss the origin of \jten\ in relation to the former generation of massive stars in the Carina Nebula and its possible evolutionary state. We interpret our findings in the light of the observed properties of the currently known neutron star population and compare \jten\ to other peculiar objects. The intriguing possibility of detecting \jten\ with the current and future generation of gravitational wave detectors is briefly discussed. Finally, we consider the possible number of Galactic neutron stars and the prospects of finding new thermally emitting sources with the \eROS\ mission \citep{pre10}, planned to be launched in 2013.
\subsection{Summary of results\label{sec_discResults}}
We have presented the results of the first dedicated observational campaign designed to investigate the properties of the isolated neutron star \jtenfull. Our analysis is based on new observations obtained with the \xmm\ Observatory and the ESO Very Large Telescope, as well as publicly available $\gamma$-ray data from the Fermi Space Telescope and the AGILE Mission. Our observational campaign has confirmed expectations based on a previous analysis of archival data and revealed a unique and peculiar object.

The new \xmm\ observation alone collected a factor of two more source counts than all previous X-ray observations (that serendipitously detected the neutron star in the past 12 years) did collectively. The higher quality statistics and higher time resolution of the pn camera in small window mode, has permitted the measurement of the period of the very likely neutron star spin, at an unexpectedly rapid rotation of $P\sim18.6$\,ms. The detection of the periodicity, corresponding to a pulsed fraction of $p_{\rm f}\sim14\%$, has a chance of $1$ in $16,000$ of being spurious. No other periodicity is significantly detected ($>1\sigma$) across a very broad period range, $P=0.0114-10000$\,s in pn data. The analysis of the three EPIC cameras together more tightly constrained the periodicities with $p_{\rm f}>10\%$ ($3\sigma$) and $P>0.6$\,s ($0.3-2.5$\,keV), thus considerably improving previous limits in this frequency range. 

The \xmm\ observation also confirmed the purely thermal nature of the source. The energy distribution is most accurately decribed by a simple blackbody of temperature $kT_\infty\sim135$\,eV, with significant deviations (absorption features) around energies 0.6\,keV and 1.35\,keV. While the first feature is more likely related to a local oxygen overabundance in the Carina Nebula or along the line-of-sight, the second one can only be accounted for by an additional spectral component, modelled as a Gaussian line in absorption. 
The source emission is very soft with practically no counts above 2\,keV; in contrast to standard rotation-powered pulsars and magnetars, a non-thermal component extending towards higher X-ray energies is detected insignificantly ($<3\sigma$) at flux levels higher than $1\%$ of the source luminosity ($L_{\rm X}=4.7^{+1.3}_{-0.9}\times10^{32}$\,erg\,s$^{-1}$, $0.1-12$\,keV). Similarly, we found no evidence of a two-temperature spectral model, as is often the case for thermally emitting radio-quiet neutron stars located near the centre of supernova remnants \citep[CCOs; see][for a review]{luc08}, some transient AXPs, or Calvera (Sect.~\ref{sec_discLinks}). 

The new \eso\ observations set an even deeper limit (relative to previous optical follow-up) to the brightness of the optical counterpart of \jten, $m_{\rm V}>27$ ($2\sigma$); the optical flux is therefore at least 6300 times fainter than the X-ray emission of the neutron star, confirming the compact nature of the source. Although a diffuse H$\alpha$ excess is present near the position of the X-ray source, similar background emission in other parts of the nebula indicates that it is unlikely to be related to the neutron star.

Our analysis of public $\gamma$-ray data has revealed no significant detection. Given the still unknown and unconstrained period derivative of the pulsar (see our discussion below), it is difficult to say whether any should be expected. By analogy with other INSs, it is likely that the pulsar spin-down luminosity is too small, given its estimated distance, to observe $\gamma$-ray emission with the present sensitivity.
\subsection{Former generation of massive stars in the Carina Nebula\label{sec_discCarina}}
The Great Nebula in Carina is an ensemble of young and rich stellar clusters, highly structured dust clouds and pillars, cavities, and bubbles created by the dramatic interaction of ionisation fronts with scattered molecular material \citep[see][for reviews]{smi08,tow11a}. Star formation is currently underway, as revealed by the presence of e.g. pre-main-sequence objects, microjets as well as many candidate young stars detected in the infrared \citep{smi10a,smi10b}. Moreover, populations of stars with ages of from 5 to 10\,Myr, as in Tr\,15 \citep{wan11}, support evidence that star formation has been ongoing for the past several million years -- i.e., a sufficiently long time for the most massive objects belonging to the former stellar generation in the complex to have already ended their lives in core-collapse events.

Although for instance \citet{smi06b} and \citet{smi08} argue that past supernovae are not needed on energetic grounds alone to create the observed cavities and superbubbles, there is growing evidence that at least one explosion has occurred in Carina: ($i$) the detection of an unusual and broad emission feature at 22\,$\mu$m, related with recently synthesized dust, which has only been observed in the Cassiopeia~A supernova remnant and in starburst galaxies \citep{cha00}; ($ii$) enhanced abundances of iron and silicon in the southern part of the nebula relative to other regions, which indicate that the metal enrichment originated in a supernova rather than winds of evolved stars \citep{ham07,ezo09}; ($iii$) the presence of high-velocity expanding structures, which are predominantly located in front of main sources of ionisation, such as the stellar cluster Tr\,16 \citep{wal07}; ($iv$) the evidence of charge exchange as the origin of Carina's diffuse soft X-ray emission, since a rarefied hot plasma resulting from an old supernova is a necessary condition for generating such a process \citep{tow11b}. 

The discovery of \jten\ provides further support for the past supernova hypothesis \citep{ham09,tow11b}. The column density derived from the X-ray spectral fits is indeed consistent with the source being located in the Nebula; its softness excludes an extragalactic object, given the Galactic absorption along the line-of-sight ($\nh=1.35\times10^{22}$\,cm$^{-2}$; \citealt{dic90}). 
\citet{tow11b} suggested that \jten\ might have been the first discovered neutron star and that others are still hidden in the Nebula, proposing an additional six X-ray sources without any evident optical or infrared counterparts as possible INS candidates (see their Table~2). In practice, however, the faintness of these candidates (observed fluxes of $f_{\rm X}\sim10^{-15}-10^{-16}$\,erg\,s$^{-1}$\,cm$^{-2}$ in the $0.5-8$\,keV energy band; \citealt{bro11}) requires unrealistically deep follow-up optical observations in order to confirm their tentative compact nature: observed upper limits in range $m_{\rm B}=29-30$ are needed to reach an X-ray-to-optical flux ratio of 100, which is still not high enough to discard either cataclysmic variables or background active galactic nuclei.

If \jten\ is indeed a remnant of a former population of massive stars in Carina, the source is still somewhat young; younger than several $10^6$\,yr, considering typical main-sequence phases of $0.5-5$\,Myr.
\subsection{Possible nature and relation to other neutron stars\label{sec_discLinks}}
The new results of our observing campaign targeting \jten\ can be used to understand how peculiar groups of isolated neutron stars relate to each other, as well as to the bulk of the normal radio pulsar population. We therefore compare the spectral and timing properties of possible ``missing links'' -- \jten, \rrat, and Calvera -- with those of other thermally emitting INSs, namely the \msev, magnetars, CCOs, and millisecond pulsars (MSPs). We first briefly review the properties of CCOs and MSPs in the next subsection (for magnetars, the \msev, Calvera and RRATs, we refer to Sect.~\ref{sec_intro}, and references therein). In the discussion, we assume that the pulsar spin down is due to magnetic dipole braking, i.e. dipolar magnetic field $B_{\rm dip}=3.2\times10^{19}(P\dot{P})^{1/2}$\,G, spin-down luminosity $\dot{E}=4.5\times10^{46}(\dot{P}P^{-3})$\,erg\,s$^{-1}$, and $\tau_{\rm ch}=P/(2\dot{P})$\,s, where $\tau_{\rm ch}$ is the pulsar characteristic time.
\subsubsection{Properties of thermally emitting INSs}
Millisecond pulsars are old and low-magnetized ($\tau_{\rm ch}=0.1-10$\,Gyr and $B_{\rm dip}\lesssim10^{10}$\,G) neutron stars, mostly observed in both binary systems and globular clusters. Accretion in a binary system is thought to be responsible for ``recycling'' these old neutron stars, first by spinning up the pulsar to millisecond periods so radio emission is turned on again, second by heating areas of the surface -- polar caps or hot spots -- which can then emit X-rays. The bulk of the X-ray radiation is usually thermal and originates from the hot spots. Power-law tails, or purely non-thermal X-ray emission, are also often observed in these objects. Interestingly, \fermi\ has also revealed that MSPs are a major contributor to the Galactic $\gamma$-ray source population, providing insights into the magnetospheric emission processes \citep[e.g.][]{abd09b,abd10}. 
\begin{figure*}[t]
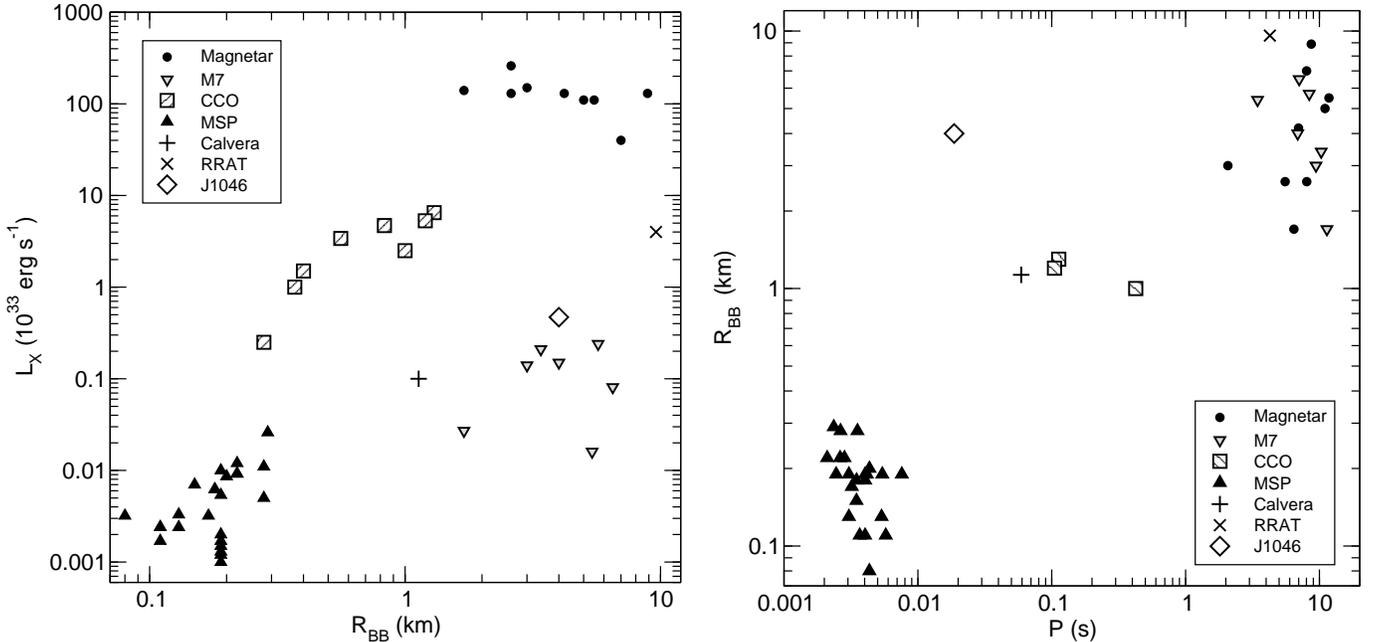

\begin{center}
\includegraphics*[width=0.49\textwidth]{fig8a.eps}\hspace{0.3cm}
\includegraphics*[width=0.47\textwidth]{fig8b.eps}
\end{center}
\caption{Blackbody luminosity versus emission radius (left) and emission radius \textit{vs.} neutron star spin period (right) for several populations of thermally emitting INSs (see text).}\label{fig_thermalINS}
\end{figure*}

Of the seven confirmed CCOs, at least three fit the ``anti-magnetar'' scenario \citep{hal10,got10a}, in which these objects are young (ages $10^3-10^4$\,yr), low-magnetized ($B\lesssim$~few~$10^{11}$\,G) neutron stars, born rotating at a period similar to its present value (of hundreds of milliseconds). Spin-down values and upper limits are very low, confining anti-magnetars to a ``transition'' region of the $P-\dot{P}$ diagram, in-between those occupied by normal rotation-powered pulsars and MSPs. 
The energy distribution of CCOs is generally described by a double blackbody model, where the emission radii (as seen by a distant observer) are typically smaller than a few km. The double blackbody, unusual pulse profiles, and strongly pulsed fractions have been interpreted in the framework of a model, where a pair of thermal, antipodal hot spots of different sizes and temperatures are present at the neutron star surface \citep[see][and the case of the CCO \object{RX~J0822-4300} in the supernova remnant Puppis~A]{got10b}. This temperature anisotropy can be understood as the result of a large difference in intensity between crustal and dipolar components of the magnetic field of the neutron star \citep{tur11,sha11}. As a CCO ages, its host supernova remnant is expected to dissipate, whereas the pulsar does not undergo considerable spin down. If standard cooling is at work \citep[e.g.][]{pag98}, the neutron star can still be observed as a thermal source for the subsequent several Myrs. However, intriguing observational evidence suggests that the CCO \object{CXOU~J232327.9+584842} in the supernova remnant Cassiopeia~A might have experienced enhanced cooling in the past 10\,yr \citep{hei10}, which has been interpreted as evidence of the onset of neutron superfluidity in the core of this young neutron star (\citealt{pag11,sht11}; alternative interpretations exist, e.g. \citealt{bla11}). That no ``orphan'' CCO (i.e. without a supernova remnant) is recognized among the X-ray emitting INSs observed in the solar vicinity, whereas they are expected to represent about one-third of the neutron star population \citep{hal10}, might indeed suggest that CCOs cool down faster than expected, also perhaps due to accreted light-element envelopes once the neutron star is in the photon-cooling stage (age $\gtrsim10^5$\,yr; \citealt{yak04}).
\subsubsection{Relations to peculiar objects and INS populations\label{sec_discRelations}}
In the following discussion, we consider the thermal X-ray luminosities ($L_{\rm X}$), emission radii ($R_{\rm bb}$), and blackbody temperatures ($kT$) reported in \citet[][]{dur06}, \citet[][]{mer08}, \citet[][for AXPs]{ngk11}, \citet[][for the \msev]{kap08a}, \citet[][and references therein; for CCOs]{hal10}, \citet[][for the X-ray properties of MSPs in 47\,Tucanae and M28]{bog06,bog11}, \citet[][for Calvera]{zan11}, and \citet[][for \rrat]{lau07}. We note that the distance to Calvera is not well-constrained; its spectral properties are hence normalised to a distance of 1\,kpc. For \jten, we use the corresponding values of $L_{\rm X}=5\times10^{32}$\,erg\,s$^{-1}$, $R_{\rm bb}=4$\,km and $kT=135$\,eV for a neutron star located at $d=2.3$\,kpc (Sect.~\ref{sec_spectra}). Other pulsar properties are taken from the ATNF Pulsar Catalogue \citep{man05}\footnote{ http://www.atnf.csiro.au/people/pulsar/psrcat}.

In Fig.~\ref{fig_thermalINS}, we compare both the X-ray luminosities of these neutron stars as a function of emission radius and the emission radius as a function of rotation period. The plots show that the four main groups -- magnetars, CCOs, the \msev, and MSPs -- occupy distinct regions of these diagrams. The $L_{\rm X}-R_{\rm bb}$ diagram (Fig.~\ref{fig_thermalINS}; left) shows that the spectral properties of \jten\ are remarkably similar to those of the \msev. This resemblance is strenghtened by the presence of a broad line in absorption in the spectrum of the source (Sect.~\ref{sec_discLine}). On the other hand, the diagram on the right in Fig.~\ref{fig_thermalINS} shows a completely different picture: the detection of \jten's very rapid rotation is puzzling and at odds with its purely thermal, \msev-like, spectrum. While all \msev\ and the RRAT have spin periods of a few seconds, which are in the same range as those observed in magnetars, \jten\ is found to be isolated in a region devoid of any other thermally emitting INS. 

In general, similar rapidly spinning neutron stars are expected to be either young and energetic rotation-powered pulsars (with $B_{\rm dip}\sim10^{12}$\,G, $\dot{E}\gtrsim10^{36}$\,erg\,s$^{-1}$, and $\tau_{\rm ch}\sim10^3-10^4$\,yr) or, conversely, old recycled neutron stars with low magnetic fields and spin-down luminosities, and old characteristic ages.
In the absence of any constraint on \jten's spin down, a rotation-powered pulsar is a viable, although unlikely, scenario. The absence of a supernova remnant or a pulsar wind nebula is indicative of an age older than a few $10^4$\,yr and a dipolar magnetic field weaker than $B_{\rm dip}\sim7\times10^{11}$\,G. However, the unusually soft and thermal energy distribution of the source, its lack of magnetospheric activity, and emission at $\gamma$-rays are in contrast to other rotation-powered pulsars of similar timing properties. 

The likely presence of \jten\ in a star-forming region (Sect.~\ref{sec_discCarina}) supports the idea that the neutron star is not old, i.e. it was most probably not (fully) recycled in a binary system with a low-mass companion that was either evaporated or completely accreted (which only happens on timescales of $10^8$\,yr). If the rapid rotation of the source were the result of spin up in a binary system, then its companion should have been another massive star. In these systems, a shorter period ($10^6-10^7$\,yr) of mass transfer takes place until the second supernova event disrupts the system. \jten\ would then be the first, ``mildly-recycled'' pulsar ejected from the system, with a spin period and a magnetic field in-between those of normal rotation-powered pulsars and old MSPs \citep[see e.g.][]{lor04,bel10}. Its former companion, which is expected to be a young pulsar, could thus still be hidden in the Carina Nebula. We note however that the luminosity and spectral properties of \jten\ are still atypical relative to those of recycled objects, as illustrated by the diagrams in Fig.~\ref{fig_thermalINS}.

Another intriguing possibility is that of an old CCO, as discussed in the case of Calvera \citep{rut08,zan11}. 
In terms of its spectral properties, Calvera somewhat differs from \jten, in that it shows clear evidence of a two-temperature thermal model and a non-thermal component extending towards higher energies cannot be excluded below $\sim10\%$ of the source luminosity. As discussed by \citeauthor{zan11}, the thermal components possibly originate in two different small spots on the stellar surface, a scenario that is consistent with Calvera's measurement of a relatively large pulsed fraction. These spectral characteristics are those expected from an older (i.e. colder and less luminous) and orphan CCO, as also suggested by the absence of a supernova remnant\footnote{As pointed out by \citet{hal11} and in spite of its spectral properties, the upper limit to Calvera's spin down, derived from X-ray timing alone, does not exclude a normal rotation-powered pulsar.}. Calvera, for a distance of $d=1$\,kpc, indeed clusters together with the three CCOs with known periods in the $R_{\rm bb}-P$ diagram of Fig.~\ref{fig_thermalINS}. Keeping in mind the low statistics and given the still poorly known overall characteristics of the population of Galactic anti-magnetars, the possibility of an old CCO remains open for \jten, although we note that the overall spectral properties of the source do not match those expected for this scenario (e.g. double thermal component, size of emission radius, and pulsed fraction).

To more clearly understand the nature of this INS, further X-ray observations would be extremely helpful, particularly given our ability to constrain the neutron star spin-down rate. For this matter, we have been granted new \xmm\ observations for the next period of observation (AO11). With a second epoch, we can constrain spin-down rates of higher than $\dot{P}\gtrsim2.5\times10^{-16}$\,s\,s$^{-1}$ and magnetic fields of $B_{\rm dip}\gtrsim6\times10^{10}$\,G ($2\sigma$), assuming standard magnetic dipole braking. The corresponding lower limit to $B_{\rm dip}$ is $\gtrsim3.4\sigma$ below the mean field strength of normal radio pulsars \citep{fau06} and provides sufficient evidence to associate \jten\ with either peculiar, low-magnetized objects, such as an old CCO and a mildly-recycled pulsar, or the normal rotation-powered population.
\subsubsection{Spectral line in absorption\label{sec_discLine}}
New \xmm\ data on \jten\ may also provide additional evidence of the spectral line at $\epsilon=1.35$\,keV (Sect.~\ref{sec_spectra}). 
Similar spectral features in absorption (in a couple of cases, also harmonically spaced) are seen in the spectra of several thermally emitting INSs: the CCO \object{1E~1207.4-5209} \citep{san02,mor05}; the rotating radio transient \rrat\ \citep{lau07}; six of the \msev\ \citep{hab07}; possibly\footnote{As noted by \citeauthor{zan11}, the contributions of the two thermal components in the spectra of Calvera intersect near $0.65$\,keV, which is probably why the spectral fit can be accommodated by introducing an additional feature -- either a Gaussian line or an edge -- around this energy.} Calvera \citep{zan11} and now \jten; these have typically been modelled by broad Gaussian absorption lines added to a simple blackbody continuum (see, however, the case of \magonethr; \citealt{hambaryan11}). In the case of the \msev, the broad features have complex and phase-dependent shapes and the line depths often exceed 50\%. The presence of narrow absorption features, possibly related to either interstellar or circumstellar highly ionised oxygen, was also reported in the co-added RGS spectra of three of the \msev\ \citep{hambaryan09,hoh12}. 

The spectral absorption features are generally understood in terms of the neutron star magnetic field, although its interpretation is not unique. They can be related to cyclotron transitions of either protons ($B_{\rm cyc}\sim10^{13}-10^{14}$\,G) or electrons ($B_{\rm cyc}\sim10^{10}-10^{11}$\,G); an alternative explanation would rest upon atomic transitions in the outermost layers of the neutron star \citep{lai01a}.

If the line at energy $\epsilon=1.35$\,keV detected in the spectrum of \jten\ is interpreted as electron cyclotron absorption similar to the case of the CCO \object{1E~1207.4-5209}, then the magnetic field of the source is expected to be of the order of $B_{\rm cyc}=1.5\times10^{11}$\,G. This estimate is based on the assumption that the line is the fundamental and that the gravitational redshift of the neutron star is $z_{\rm g}=0.3$, according to the relation $E_{\rm cyc}=1.16(B/10^{11}\textrm{\,G})(1+z_{\rm g})^{-1}$. Alternatively, the feature could be the only first harmonic of a fundamental at an energy $\epsilon\sim0.65-0.7$\,keV, whose detection is unfortunately affected by our interpretation of a local oxygen overabundance in the Carina Nebula. In this case, the magnetic field would be lower, $B=7\times10^{10}$\,G. These estimates can be confronted with the upper limits derived from the pulsar spin down.
\subsection{Future prospects\label{sec_discProspects}}
The knowledge of the pulsar spin down (and its associated quantities $B_{\rm dip}$, $\tau_{\rm ch}$, $\dot{E}$) is crucial to help us characterise the source \jten. Furthermore, a constraint on the period derivative significantly narrows the parameter space of searches for a pulsed $\gamma$-ray signal or even for gravitational waves from the neutron star.

Gravitational wave observatories have the potential to probe several aspects of neutron star physics, in particular providing constraints on the state of matter at extreme densities \citep[see e.g.][for a review]{and11}. 
Although a detection has not yet been reported, the observatories have provided a wealth of interesting information on the Galactic neutron star population \citep[e.g.][]{abb08,abb10,aba10a,aba10b,aba11a,aba11b,aba12,acc10}. The second generation of gravitational wave detectors (e.g. Advanced LIGO) might allow the first detection \citep{and11}.

Rapidly-spinning and young neutron stars in our Galaxy are the best targets to search for a continuous gravitational wave signal with ground-based interferometric observatories. 
The rapid rotation of \jten\ implies that it emits continuous gravitational waves at a frequency $f\sim110$\,Hz, where LIGO's $S/N$ is the highest \citep[see][their Fig.~4]{abb10}. In principle, a spin-down rate as low as $\dot{P}\sim8\times10^{-16}$\,s\,s$^{-1}$, which may be detected with the new \xmm\ observation planned in AO11, places \jten\ within the sensitivity of the LIGO's S5 Scientific Run, considering the whole data set of integration. This estimate, of course, simplistically assumes a 100\% conversion of spin-down energy into gravitational waves, regardless of how large a neutron star asymmetry is required to power such emission. 
In any case, its rapid rotation, quite young age, and relative proximity make this neutron star in the Carina Nebula an interesting prospect for gravitational wave detection with Advanced LIGO (B.~Owen, private communication).

In spite of many searches and interesting case studies, no other thermally emitting INS, presenting exactly the same characteristics as the \msev, has been identified outside the solar vicinity to date (see also the case of the high-$B$ radio pulsar \object{PSR~J0726-2612}, which is another likely product of the Gould Belt; \citealt{spe11}). 
It is therefore very important for population studies to understand why there are so many thermally emitting sources with similar periods (and presumably ages and magnetic fields) in such a small volume. 
Is this an anomaly caused by the Sun's current location close to regions of active stellar formation of the Gould Belt or does it really mean that radio surveys do miss a large population of INSs, at least as large as that of normal radio pulsars? To answer these questions, investigations at fainter fluxes as well as population modelling in the Galactic scale are needed. 

\eROS\ (Extended R\"ontgen Survey with an Imaging Telescope Array\footnote{ http://www.mpe.mpg.de/projects.html\#erosita}) will extend the \ros\ All-Sky Survey \citep{vog99b} towards higher energies, with unprecedented spectral and angular resolution. 
The effective area of \eROS\ is twice that of one \xmm\ telescope at energies below 2\,keV and its sensitivity during the planned all-sky survey will be approximately 30 times that of \ros. These properties combined will allow an estimated number of from about 100 to 200 new X-ray thermally emitting INSs to be discovered, after four years of the \eROS\ survey.
The above figures are derived using the population synthesis model described in \citet{pir09c}\footnote{Available at http://scd-theses.u-strasbg.fr/1876}. 
Expectations for the same population based on the model of Popov, Boldin, and collaborators predict a more modest number, of $\sim30$ new discoveries after the planned years of survey. Although many of the assumptions considered in our population synthesis model differ from those of Popov et al., preventing a proper comparison, we estimate that the likely parameters controlling the discrepancy are the adopted cooling rates and, to a lower extent, the model of the interstellar medium distribution.
\section{Conclusions\label{sec_summary}}
Our observational campaign on \jtenfull\ has confirmed previous expectations and revealed a unique object. Most notably, our deep \xmm\ observation has allowed us to unveil the rotation of the neutron star: a very fast spin period of 19\,ms is surprisingly at odds with \jten's purely thermal energy distribution, which is reminiscent of those of the Magnificent Seven. Spectral features in absorption have also been positively identified. The nature of the source remains open to different possible interpretations, a relation to the still poorly known class of Galactic anti-magnetars being a favoured scenario. New \xmm\ data, granted for the next cycle of observations (AO11), will greatly improve the current observational picture of this source, and enable us to tightly constrain the pulsar spin down. Dedicated radio observations would be similarly invaluable to help us unveil any truely radio-quiet nature.
\begin{acknowledgements}
The authors would like to express their thanks to R.~Rutledge and B.~Owen for fruitful discussions and suggestions. We also thank the \xmm\ Helpdesk, in particular M. Ehle and R~Gonzalez-Riestra, for extensive discussions during the preparation of the phase~II of the AO9/AO11 observations. 
The work of A.M.P. is supported by the Deutsche Forschungsgemeinschaft (grant PI~983/1-1). A.M.P. gratefully acknowledges support from the Alexander von Humboldt Foundation (Fellowship for Postdoctoral Researchers) and FAPESP, Brazil (grant 2009\_18499-6). 
\end{acknowledgements}
\bibliographystyle{aa}
\bibliography{ins}

\hyphenation{Post-Script Sprin-ger}
\begin{thebibliography}{119}
\expandafter\ifx\csname natexlab\endcsname\relax\def\natexlab#1{#1}\fi

\bibitem[{{Abadie} {et~al.}(2011{\natexlab{a}}){Abadie}, {Abbott}, {Abbott},
  {Abbott}, {Abernathy}, {Accadia}, {Acernese}, {Adams}, {Adhikari}, {Affeldt},
  \& et~al.}]{aba11b}
{Abadie}, J., {Abbott}, B.~P., {Abbott}, R., {et~al.} 2011{\natexlab{a}}, ArXiv
  e-prints, astro-ph/1112.5004

\bibitem[{{Abadie} {et~al.}(2012){Abadie}, {Abbott}, {Abbott}, {Abbott},
  {Abernathy}, {Accadia}, {Acernese}, {Adams}, {Adhikari}, {Affeldt}, \&
  et~al.}]{aba12}
{Abadie}, J., {Abbott}, B.~P., {Abbott}, R., {et~al.} 2012, \prd, 85, 022001

\bibitem[{{Abadie} {et~al.}(2010{\natexlab{a}}){Abadie}, {Abbott}, {Abbott},
  {Abernathy}, {Accadia}, {Acernese}, {Adams}, {Adhikari}, {Ajith}, {Allen}, \&
  et~al.}]{aba10b}
{Abadie}, J., {Abbott}, B.~P., {Abbott}, R., {et~al.} 2010{\natexlab{a}}, \prd,
  82, 102001

\bibitem[{{Abadie} {et~al.}(2011{\natexlab{b}}){Abadie}, {Abbott}, {Abbott},
  {Abernathy}, {Accadia}, {Acernese}, {Adams}, {Adhikari}, {Ajith}, {Allen}, \&
  et~al.}]{aba11a}
{Abadie}, J., {Abbott}, B.~P., {Abbott}, R., {et~al.} 2011{\natexlab{b}},
  Physical Review Letters, 107, A261102

\bibitem[{{Abadie} {et~al.}(2010{\natexlab{b}}){Abadie}, {Abbott}, {Abbott},
  {Abernathy}, {Adams}, {Adhikari}, {Ajith}, {Allen}, {Allen}, {Amador Ceron},
  \& et~al.}]{aba10a}
{Abadie}, J., {Abbott}, B.~P., {Abbott}, R., {et~al.} 2010{\natexlab{b}}, \apj,
  722, 1504

\bibitem[{{Abbott} {et~al.}(2008){Abbott}, {Abbott}, {Adhikari}, {Ajith},
  {Allen}, {Allen}, {Amin}, {Anderson}, {Anderson}, {Arain}, \& et~al.}]{abb08}
{Abbott}, B., {Abbott}, R., {Adhikari}, R., {et~al.} 2008, \apjl, 683, L45

\bibitem[{{Abbott} {et~al.}(2010){Abbott}, {Abbott}, {Acernese}, {Adhikari},
  {Ajith}, {Allen}, {Allen}, {Alshourbagy}, {Amin}, {Anderson}, \&
  et~al.}]{abb10}
{Abbott}, B.~P., {Abbott}, R., {Acernese}, F., {et~al.} 2010, \apj, 713, 671

\bibitem[{{Abdo} {et~al.}(2009){Abdo}, {Ackermann}, {Ajello}, {Atwood},
  {Axelsson}, {Baldini}, {Ballet}, {Barbiellini}, {Bastieri}, {Baughman},
  {Bechtol}, {Bellazzini}, {Berenji}, {Blandford}, {Bloom}, {Bonamente},
  {Borgland}, {Bregeon}, {Brez}, {Brigida}, {Bruel}, {Burnett}, {Caliandro},
  {Cameron}, {Caraveo}, {Casandjian}, {Cecchi}, {{\c C}elik}, {Charles},
  {Chaty}, {Chekhtman}, {Cheung}, {Chiang}, {Ciprini}, {Claus}, {Cohen-Tanugi},
  {Conrad}, {Cutini}, {Dermer}, {de Palma}, {Digel}, {Dormody}, {do Couto e
  Silva}, {Drell}, {Dubois}, {Dumora}, {Farnier}, {Favuzzi}, {Fegan}, {Focke},
  {Frailis}, {Fukazawa}, {Fusco}, {Gargano}, {Gasparrini}, {Gehrels},
  {Germani}, {Giebels}, {Giglietto}, {Giordano}, {Glanzman}, {Godfrey},
  {Grenier}, {Grove}, {Guillemot}, {Guiriec}, {Hanabata}, {Harding},
  {Hayashida}, {Hays}, {Horan}, {Hughes}, {J{\'o}hannesson}, {Johnson},
  {Johnson}, {Johnson}, {Johnson}, {Kamae}, {Katagiri}, {Kawai}, {Kerr},
  {Kn{\"o}dlseder}, {Kuehn}, {Kuss}, {Lande}, {Latronico}, {Lemoine-Goumard},
  {Longo}, {Loparco}, {Lott}, {Lovellette}, {Lubrano}, {Makeev}, {Mazziotta},
  {McConville}, {McEnery}, {Meurer}, {Michelson}, {Mitthumsiri}, {Mizuno},
  {Moiseev}, {Monte}, {Monzani}, {Morselli}, {Moskalenko}, {Murgia}, {Nolan},
  {Norris}, {Nuss}, {Ohsugi}, {Omodei}, {Orlando}, {Ormes}, {Paneque},
  {Panetta}, {Parent}, {Pelassa}, {Pepe}, {Pierbattista}, {Piron}, {Porter},
  {Rain{\`o}}, {Rando}, {Razzano}, {Rea}, {Reimer}, {Reimer}, {Reposeur},
  {Ritz}, {Rochester}, {Rodriguez}, {Romani}, {Roth}, {Ryde}, {Sadrozinski},
  {Sanchez}, {Sander}, {Saz Parkinson}, {Sgr{\`o}}, {Smith}, {Smith},
  {Spandre}, {Spinelli}, {Starck}, {Strickman}, {Suson}, {Tajima}, {Takahashi},
  {Tanaka}, {Thayer}, {Thayer}, {Thompson}, {Tibaldo}, {Torres}, {Tosti},
  {Tramacere}, {Uchiyama}, {Usher}, {Vasileiou}, {Vilchez}, {Vitale}, {Wang},
  {Webb}, {Winer}, {Wood}, {Ylinen}, \& {Ziegler}}]{abd09b}
{Abdo}, A.~A., {Ackermann}, M., {Ajello}, M., {et~al.} 2009, Science, 325, 845

\bibitem[{{Abdo} {et~al.}(2010){Abdo}, {Ackermann}, {Ajello}, {Baldini},
  {Ballet}, {Barbiellini}, {Bastieri}, {Bellazzini}, {Blandford}, {Bloom},
  {Bonamente}, {Borgland}, {Bouvier}, {Brandt}, {Bregeon}, {Brigida}, {Bruel},
  {Buehler}, {Buson}, {Caliandro}, {Cameron}, {Caraveo}, {Carrigan},
  {Casandjian}, {Charles}, {Chaty}, {Chekhtman}, {Cheung}, {Chiang}, {Ciprini},
  {Claus}, {Cohen-Tanugi}, {Conrad}, {Decesar}, {Dermer}, {de Palma}, {Digel},
  {Silva}, {Drell}, {Dubois}, {Dumora}, {Favuzzi}, {Fortin}, {Frailis},
  {Fukazawa}, {Fusco}, {Gargano}, {Gasparrini}, {Gehrels}, {Germani},
  {Giglietto}, {Giordano}, {Glanzman}, {Godfrey}, {Grenier}, {Grondin},
  {Grove}, {Guillemot}, {Guiriec}, {Hadasch}, {Harding}, {Hays}, {Jean},
  {J{\'o}hannesson}, {Johnson}, {Johnson}, {Kamae}, {Katagiri}, {Kataoka},
  {Kerr}, {Kn{\"o}dlseder}, {Kuss}, {Lande}, {Latronico}, {Lee},
  {Lemoine-Goumard}, {Llena Garde}, {Longo}, {Loparco}, {Lovellette},
  {Lubrano}, {Makeev}, {Mazziotta}, {Michelson}, {Mitthumsiri}, {Mizuno},
  {Monte}, {Monzani}, {Morselli}, {Moskalenko}, {Murgia}, {Naumann-Godo},
  {Nolan}, {Norris}, {Nuss}, {Ohsugi}, {Omodei}, {Orlando}, {Ormes},
  {Pancrazi}, {Parent}, {Pepe}, {Pesce-Rollins}, {Piron}, {Porter},
  {Rain{\`o}}, {Rando}, {Reimer}, {Reimer}, {Reposeur}, {Ripken}, {Romani},
  {Roth}, {Sadrozinski}, {Saz Parkinson}, {Sgr{\`o}}, {Siskind}, {Smith},
  {Spinelli}, {Strickman}, {Suson}, {Takahashi}, {Takahashi}, {Tanaka},
  {Thayer}, {Thayer}, {Tibaldo}, {Torres}, {Tosti}, {Tramacere}, {Uchiyama},
  {Usher}, {Vasileiou}, {Venter}, {Vilchez}, {Vitale}, {Waite}, {Wang}, {Webb},
  {Winer}, {Yang}, {Ylinen}, \& {Ziegler}}]{abd10}
{Abdo}, A.~A., {Ackermann}, M., {Ajello}, M., {et~al.} 2010, \aap, 524, A75

\bibitem[{{Accadia} {et~al.}(2010){Accadia}, {Swinkels}, \& {the VIRGO
  Collaboration}}]{acc10}
{Accadia}, T., {Swinkels}, B.~L., \& {the VIRGO Collaboration}. 2010, Classical
  and Quantum Gravity, 27, 084002

\bibitem[{{Ag{\"u}eros} {et~al.}(2006){Ag{\"u}eros}, {Anderson}, {Margon},
  {Posselt}, {Haberl}, {Voges}, {Annis}, {Schneider}, \& {Brinkmann}}]{agu06}
{Ag{\"u}eros}, M.~A., {Anderson}, S.~F., {Margon}, B., {et~al.} 2006, \aj, 131,
  1740

\bibitem[{{Ag{\"u}eros} {et~al.}(2011){Ag{\"u}eros}, {Posselt}, {Anderson},
  {Rosenfield}, {Haberl}, {Homer}, {Margon}, {Newsom}, \& {Voges}}]{agu11}
{Ag{\"u}eros}, M.~A., {Posselt}, B., {Anderson}, S.~F., {et~al.} 2011, \aj,
  141, 176

\bibitem[{{Aguilera} {et~al.}(2008){Aguilera}, {Pons}, \& {Miralles}}]{agu08}
{Aguilera}, D.~N., {Pons}, J.~A., \& {Miralles}, J.~A. 2008, \apjl, 673, L167

\bibitem[{{Anders} \& {Grevesse}(1989)}]{and89}
{Anders}, E. \& {Grevesse}, N. 1989, \gca, 53, 197

\bibitem[{{Andersson} {et~al.}(2011){Andersson}, {Ferrari}, {Jones},
  {Kokkotas}, {Krishnan}, {Read}, {Rezzolla}, \& {Zink}}]{and11}
{Andersson}, N., {Ferrari}, V., {Jones}, D.~I., {et~al.} 2011, General
  Relativity and Gravitation, 43, 409

\bibitem[{{Appenzeller} {et~al.}(1998){Appenzeller}, {Fricke}, {F{\"u}rtig},
  {G{\"a}ssler}, {H{\"a}fner}, {Harke}, {Hess}, {Hummel}, {J{\"u}rgens},
  {Kudritzki}, {Mantel}, {Meisl}, {Muschielok}, {Nicklas}, {Rupprecht},
  {Seifert}, {Stahl}, {Szeifert}, \& {Tarantik}}]{app98}
{Appenzeller}, I., {Fricke}, K., {F{\"u}rtig}, W., {et~al.} 1998, The
  Messenger, 94, 1

\bibitem[{{Atwood} {et~al.}(2009){Atwood}, {Abdo}, {Ackermann}, {Althouse},
  {Anderson}, {Axelsson}, {Baldini}, {Ballet}, {Band}, {Barbiellini}, \&
  et~al.}]{atw09}
{Atwood}, W.~B., {Abdo}, A.~A., {Ackermann}, M., {et~al.} 2009, \apj, 697, 1071

\bibitem[{{Belczynski} {et~al.}(2010){Belczynski}, {Lorimer}, {Ridley}, \&
  {Curran}}]{bel10}
{Belczynski}, K., {Lorimer}, D.~R., {Ridley}, J.~P., \& {Curran}, S.~J. 2010,
  \mnras, 407, 1245

\bibitem[{{Blaes} {et~al.}(1995){Blaes}, {Warren}, \& {Madau}}]{bla95}
{Blaes}, O., {Warren}, O., \& {Madau}, P. 1995, \apj, 454, 370

\bibitem[{{Blaschke} {et~al.}(2011){Blaschke}, {Grigorian}, {Voskresensky}, \&
  {Weber}}]{bla11}
{Blaschke}, D., {Grigorian}, H., {Voskresensky}, D.~N., \& {Weber}, F. 2011,
  ArXiv e-prints, astro-ph/1108.4125

\bibitem[{{Bogdanov} {et~al.}(2006){Bogdanov}, {Grindlay}, {Heinke}, {Camilo},
  {Freire}, \& {Becker}}]{bog06}
{Bogdanov}, S., {Grindlay}, J.~E., {Heinke}, C.~O., {et~al.} 2006, \apj, 646,
  1104

\bibitem[{{Bogdanov} {et~al.}(2011){Bogdanov}, {van den Berg}, {Servillat},
  {Heinke}, {Grindlay}, {Stairs}, {Ransom}, {Freire}, {B{\'e}gin}, \&
  {Becker}}]{bog11}
{Bogdanov}, S., {van den Berg}, M., {Servillat}, M., {et~al.} 2011, \apj, 730,
  81

\bibitem[{{Broos} {et~al.}(2011){Broos}, {Townsley}, {Feigelson}, {Getman},
  {Garmire}, {Preibisch}, {Smith}, {Babler}, {Hodgkin}, {Indebetouw}, {Irwin},
  {King}, {Lewis}, {Majewski}, {McCaughrean}, {Meade}, \& {Zinnecker}}]{bro11}
{Broos}, P.~S., {Townsley}, L.~K., {Feigelson}, E.~D., {et~al.} 2011, \apjs,
  194, 2

\bibitem[{{Buccheri} {et~al.}(1983){Buccheri}, {Bennett}, {Bignami}, {Bloemen},
  {Boriakoff}, {Caraveo}, {Hermsen}, {Kanbach}, {Manchester}, {Masnou},
  {Mayer-Hasselwander}, {Ozel}, {Paul}, {Sacco}, {Scarsi}, \& {Strong}}]{buc83}
{Buccheri}, R., {Bennett}, K., {Bignami}, G.~F., {et~al.} 1983, \aap, 128, 245

\bibitem[{{Chan} \& {Onaka}(2000)}]{cha00}
{Chan}, K.-W. \& {Onaka}, T. 2000, \apjl, 533, L33

\bibitem[{{Chatterjee} \& {Cordes}(2002)}]{cha02}
{Chatterjee}, S. \& {Cordes}, J.~M. 2002, \apj, 575, 407

\bibitem[{{Chieregato} {et~al.}(2005){Chieregato}, {Campana}, {Treves},
  {Moretti}, {Mignani}, \& {Tagliaferri}}]{chi05}
{Chieregato}, M., {Campana}, S., {Treves}, A., {et~al.} 2005, \aap, 444, 69

\bibitem[{{Cordes} {et~al.}(1993){Cordes}, {Romani}, \& {Lundgren}}]{cor93a}
{Cordes}, J.~M., {Romani}, R.~W., \& {Lundgren}, S.~C. 1993, \nat, 362, 133

\bibitem[{{de Luca}(2008)}]{luc08}
{de Luca}, A. 2008, in American Institute of Physics Conference Series, Vol.
  983, 40 Years of Pulsars: Millisecond Pulsars, Magnetars and More, ed.
  C.~{Bassa}, Z.~{Wang}, A.~{Cumming}, \& V.~M. {Kaspi}, 311--319

\bibitem[{{De Luca} {et~al.}(2005){De Luca}, {Caraveo}, {Mereghetti},
  {Negroni}, \& {Bignami}}]{luc05}
{De Luca}, A., {Caraveo}, P.~A., {Mereghetti}, S., {Negroni}, M., \& {Bignami},
  G.~F. 2005, \apj, 623, 1051

\bibitem[{{Dickey} \& {Lockman}(1990)}]{dic90}
{Dickey}, J.~M. \& {Lockman}, F.~J. 1990, \araa, 28, 215

\bibitem[{{Durant} \& {van Kerkwijk}(2006)}]{dur06}
{Durant}, M. \& {van Kerkwijk}, M.~H. 2006, \apj, 650, 1070

\bibitem[{{Ezoe} {et~al.}(2009){Ezoe}, {Hamaguchi}, {Gruendl}, {Chu}, {Petre},
  \& {Corcoran}}]{ezo09}
{Ezoe}, Y., {Hamaguchi}, K., {Gruendl}, R.~A., {et~al.} 2009, \pasj, 61, 123

\bibitem[{{Faucher-Gigu{\`e}re} \& {Kaspi}(2006)}]{fau06}
{Faucher-Gigu{\`e}re}, C.-A. \& {Kaspi}, V.~M. 2006, \apj, 643, 332

\bibitem[{{Gaensler} \& {Slane}(2006)}]{gae06}
{Gaensler}, B.~M. \& {Slane}, P.~O. 2006, \araa, 44, 17

\bibitem[{{Gotthelf} \& {Halpern}(2010)}]{got10a}
{Gotthelf}, E.~V. \& {Halpern}, J.~P. 2010, in Bulletin of the American
  Astronomical Society, Vol.~42, AAS/High Energy Astrophysics Division \#11,
  694

\bibitem[{{Gotthelf} {et~al.}(2010){Gotthelf}, {Perna}, \& {Halpern}}]{got10b}
{Gotthelf}, E.~V., {Perna}, R., \& {Halpern}, J.~P. 2010, \apj, 724, 1316

\bibitem[{{Haberl}(2007)}]{hab07}
{Haberl}, F. 2007, \apss, 308, 181

\bibitem[{{Halpern}(2011)}]{hal11}
{Halpern}, J.~P. 2011, \apjl, 736, L3+

\bibitem[{{Halpern} \& {Gotthelf}(2010)}]{hal10}
{Halpern}, J.~P. \& {Gotthelf}, E.~V. 2010, \apj, 709, 436

\bibitem[{{Hamaguchi} {et~al.}(2009){Hamaguchi}, {Corcoran}, {Ezoe},
  {Townsley}, {Broos}, {Gruendl}, {Vaidya}, {White}, {Strohmayer}, {Petre}, \&
  {Chu}}]{ham09}
{Hamaguchi}, K., {Corcoran}, M.~F., {Ezoe}, Y., {et~al.} 2009, \apjl, 695, L4

\bibitem[{{Hamaguchi} {et~al.}(2007){Hamaguchi}, {Petre}, {Matsumoto},
  {Tsujimoto}, {Holt}, {Ezoe}, {Ozawa}, {Tsuboi}, {Soong}, {Kitamoto},
  {Sekiguchi}, \& {Kokubun}}]{ham07}
{Hamaguchi}, K., {Petre}, R., {Matsumoto}, H., {et~al.} 2007, \pasj, 59, 151

\bibitem[{{Hambaryan} {et~al.}(2009){Hambaryan}, {Neuh{\"a}user}, {Haberl},
  {Hohle}, \& {Schwope}}]{hambaryan09}
{Hambaryan}, V., {Neuh{\"a}user}, R., {Haberl}, F., {Hohle}, M.~M., \&
  {Schwope}, A.~D. 2009, \aap, 497, L9

\bibitem[{{Hambaryan} {et~al.}(2011){Hambaryan}, {Suleimanov}, {Schwope},
  {Neuh{\"a}user}, {Werner}, \& {Potekhin}}]{hambaryan11}
{Hambaryan}, V., {Suleimanov}, V., {Schwope}, A.~D., {et~al.} 2011, \aap, 534,
  A74

\bibitem[{{Hamuy} {et~al.}(1994){Hamuy}, {Suntzeff}, {Heathcote}, {Walker},
  {Gigoux}, \& {Phillips}}]{ham94}
{Hamuy}, M., {Suntzeff}, N.~B., {Heathcote}, S.~R., {et~al.} 1994, \pasp, 106,
  566

\bibitem[{{Heinke} \& {Ho}(2010)}]{hei10}
{Heinke}, C.~O. \& {Ho}, W.~C.~G. 2010, \apjl, 719, L167

\bibitem[{{Heyl} \& {Kulkarni}(1998)}]{hey98}
{Heyl}, J.~S. \& {Kulkarni}, S.~R. 1998, \apjl, 506, L61

\bibitem[{{Hohle} {et~al.}(2012){Hohle}, {Haberl}, {Vink}, {de Vries}, \&
  {Neuh{\"a}user}}]{hoh12}
{Hohle}, M.~M., {Haberl}, F., {Vink}, J., {de Vries}, C.~P., \&
  {Neuh{\"a}user}, R. 2012, \mnras, 419, 1525

\bibitem[{{Jansen} {et~al.}(2001){Jansen}, {Lumb}, {Altieri}, {Clavel}, {Ehle},
  {Erd}, {Gabriel}, {Guainazzi}, {Gondoin}, {Much}, {Munoz}, {Santos},
  {Schartel}, {Texier}, \& {Vacanti}}]{jan01}
{Jansen}, F., {Lumb}, D., {Altieri}, B., {et~al.} 2001, \aap, 365, L1

\bibitem[{{Kaplan}(2008)}]{kap08a}
{Kaplan}, D.~L. 2008, in American Institute of Physics Conference Series, Vol.
  983, 40 Years of Pulsars: Millisecond Pulsars, Magnetars and More, ed.
  C.~{Bassa}, Z.~{Wang}, A.~{Cumming}, \& V.~M. {Kaspi}, 331--339

\bibitem[{{Kaplan} {et~al.}(2009){Kaplan}, {Esposito}, {Chatterjee},
  {Possenti}, {McLaughlin}, {Camilo}, {Chakrabarty}, \& {Slane}}]{kap09c}
{Kaplan}, D.~L., {Esposito}, P., {Chatterjee}, S., {et~al.} 2009, \mnras, 400,
  1445

\bibitem[{{Kaplan} {et~al.}(2011){Kaplan}, {Kamble}, {van Kerkwijk}, \&
  {Ho}}]{kap11a}
{Kaplan}, D.~L., {Kamble}, A., {van Kerkwijk}, M.~H., \& {Ho}, W.~C.~G. 2011,
  \apj, 736, 117

\bibitem[{{Kaplan} \& {van Kerkwijk}(2009)}]{kap09d}
{Kaplan}, D.~L. \& {van Kerkwijk}, M.~H. 2009, \apj, 705, 798

\bibitem[{{Kaplan} \& {van Kerkwijk}(2011)}]{kap11b}
{Kaplan}, D.~L. \& {van Kerkwijk}, M.~H. 2011, \apjl, 740, L30+

\bibitem[{{Keane} {et~al.}(2011){Keane}, {Kramer}, {Lyne}, {Stappers}, \&
  {McLaughlin}}]{kea11a}
{Keane}, E.~F., {Kramer}, M., {Lyne}, A.~G., {Stappers}, B.~W., \&
  {McLaughlin}, M.~A. 2011, \mnras, 415, 3065

\bibitem[{{Keane} {et~al.}(2010){Keane}, {Ludovici}, {Eatough}, {Kramer},
  {Lyne}, {McLaughlin}, \& {Stappers}}]{kea10a}
{Keane}, E.~F., {Ludovici}, D.~A., {Eatough}, R.~P., {et~al.} 2010, \mnras,
  401, 1057

\bibitem[{{Keane} \& {McLaughlin}(2011)}]{kea11b}
{Keane}, E.~F. \& {McLaughlin}, M.~A. 2011, Bulletin of the Astronomical
  Society of India, 39, 333

\bibitem[{{Kondratiev} {et~al.}(2009){Kondratiev}, {McLaughlin}, {Lorimer},
  {Burgay}, {Possenti}, {Turolla}, {Popov}, \& {Zane}}]{kon09}
{Kondratiev}, V.~I., {McLaughlin}, M.~A., {Lorimer}, D.~R., {et~al.} 2009,
  \apj, 702, 692

\bibitem[{{Lai}(2001)}]{lai01a}
{Lai}, D. 2001, Reviews of Modern Physics, 73, 629

\bibitem[{{Lorimer} {et~al.}(2004){Lorimer}, {McLaughlin}, {Arzoumanian},
  {Xilouris}, {Cordes}, {Lommen}, {Fruchter}, {Chandler}, \& {Backer}}]{lor04}
{Lorimer}, D.~R., {McLaughlin}, M.~A., {Arzoumanian}, Z., {et~al.} 2004,
  \mnras, 347, L21

\bibitem[{{Lyne} {et~al.}(2009){Lyne}, {McLaughlin}, {Keane}, {Kramer},
  {Espinoza}, {Stappers}, {Palliyaguru}, \& {Miller}}]{lyn09}
{Lyne}, A.~G., {McLaughlin}, M.~A., {Keane}, E.~F., {et~al.} 2009, \mnras, 400,
  1439

\bibitem[{{Manchester} {et~al.}(2005){Manchester}, {Hobbs}, {Teoh}, \&
  {Hobbs}}]{man05}
{Manchester}, R.~N., {Hobbs}, G.~B., {Teoh}, A., \& {Hobbs}, M. 2005, \aj, 129,
  1993

\bibitem[{{McLaughlin} {et~al.}(2006){McLaughlin}, {Lyne}, {Lorimer}, {Kramer},
  {Faulkner}, {Manchester}, {Cordes}, {Camilo}, {Possenti}, {Stairs}, {Hobbs},
  {D'Amico}, {Burgay}, \& {O'Brien}}]{lau06}
{McLaughlin}, M.~A., {Lyne}, A.~G., {Lorimer}, D.~R., {et~al.} 2006, \nat, 439,
  817

\bibitem[{{McLaughlin} {et~al.}(2007){McLaughlin}, {Rea}, {Gaensler},
  {Chatterjee}, {Camilo}, {Kramer}, {Lorimer}, {Lyne}, {Israel}, \&
  {Possenti}}]{lau07}
{McLaughlin}, M.~A., {Rea}, N., {Gaensler}, B.~M., {et~al.} 2007, \apj, 670,
  1307

\bibitem[{{Mereghetti}(2008)}]{mer08}
{Mereghetti}, S. 2008, \aapr, 15, 225

\bibitem[{{Mori} {et~al.}(2005){Mori}, {Chonko}, \& {Hailey}}]{mor05}
{Mori}, K., {Chonko}, J.~C., \& {Hailey}, C.~J. 2005, \apj, 631, 1082

\bibitem[{{Motch} {et~al.}(2009){Motch}, {Pires}, {Haberl}, {Schwope}, \&
  {Zavlin}}]{mot09}
{Motch}, C., {Pires}, A.~M., {Haberl}, F., {Schwope}, A., \& {Zavlin}, V.~E.
  2009, \aap, 497, 423

\bibitem[{{Motch} {et~al.}(2005){Motch}, {Sekiguchi}, {Haberl}, {Zavlin},
  {Schwope}, \& {Pakull}}]{mot05}
{Motch}, C., {Sekiguchi}, K., {Haberl}, F., {et~al.} 2005, \aap, 429, 257

\bibitem[{{Ng} {et~al.}(2011){Ng}, {Kaspi}, {Dib}, {Olausen}, {Scholz},
  {G{\"u}ver}, {{\"O}zel}, {Gavriil}, \& {Woods}}]{ngk11}
{Ng}, C.-Y., {Kaspi}, V.~M., {Dib}, R., {et~al.} 2011, \apj, 729, 131

\bibitem[{{Page}(1998)}]{pag98}
{Page}, D. 1998, in NATO ASIC Proc. 515: The Many Faces of Neutron Stars., ed.
  R.~{Buccheri}, J.~{van Paradijs}, \& A.~{Alpar}, 539--+

\bibitem[{{Page} {et~al.}(2011){Page}, {Prakash}, {Lattimer}, \&
  {Steiner}}]{pag11}
{Page}, D., {Prakash}, M., {Lattimer}, J.~M., \& {Steiner}, A.~W. 2011,
  Physical Review Letters, 106, 081101

\bibitem[{{Pavlov} {et~al.}(1995){Pavlov}, {Shibanov}, {Zavlin}, \&
  {Meyer}}]{pav95}
{Pavlov}, G.~G., {Shibanov}, Y.~A., {Zavlin}, V.~E., \& {Meyer}, R.~D. 1995, in
  The Lives of the Neutron Stars, ed. M.~A. {Alpar}, U.~{Kiziloglu}, \& J.~{van
  Paradijs}, 71--+

\bibitem[{{Pellizzoni} {et~al.}(2009{\natexlab{a}}){Pellizzoni}, {Pilia},
  {Possenti}, {Chen}, {Giuliani}, {Trois}, {Caraveo}, {Del Monte}, {Fornari},
  {Fuschino}, {Mereghetti}, {Tavani}, {Argan}, {Burgay}, {Cognard}, {Corongiu},
  {Costa}, {D'Amico}, {De Luca}, {Esposito}, {Evangelista}, {Feroci},
  {Johnston}, {Kramer}, {Longo}, {Marisaldi}, {Theureau}, {Weltevrede},
  {Barbiellini}, {Boffelli}, {Bulgarelli}, {Cattaneo}, {Cocco}, {D'Ammando},
  {DeParis}, {Di Cocco}, {Donnarumma}, {Fiorini}, {Froysland}, {Galli},
  {Gianotti}, {Labanti}, {Lapshov}, {Lazzarotto}, {Lipari}, {Mineo},
  {Morselli}, {Pacciani}, {Perotti}, {Piano}, {Picozza}, {Prest}, {Pucella},
  {Rapisarda}, {Rappoldi}, {Sabatini}, {Soffitta}, {Trifoglio}, {Vallazza},
  {Vercellone}, {Vittorini}, {Zambra}, {Zanello}, {Pittori}, {Verrecchia},
  {Preger}, {Santolamazza}, {Giommi}, {Salotti}, \& {Bignami}}]{pel09b}
{Pellizzoni}, A., {Pilia}, M., {Possenti}, A., {et~al.} 2009{\natexlab{a}},
  \apjl, 695, L115

\bibitem[{{Pellizzoni} {et~al.}(2009{\natexlab{b}}){Pellizzoni}, {Pilia},
  {Possenti}, {Fornari}, {Caraveo}, {del Monte}, {Mereghetti}, {Tavani},
  {Argan}, {Trois}, {Burgay}, {Chen}, {Cognard}, {Costa}, {D'Amico},
  {Esposito}, {Evangelista}, {Feroci}, {Fuschino}, {Giuliani}, {Halpern},
  {Hobbs}, {Hotan}, {Johnston}, {Kramer}, {Longo}, {Manchester}, {Marisaldi},
  {Palfreyman}, {Weltevrede}, {Barbiellini}, {Boffelli}, {Bulgarelli},
  {Cattaneo}, {Cocco}, {D'Ammando}, {DeParis}, {Di Cocco}, {Donnarumma},
  {Fiorini}, {Froysland}, {Galli}, {Gianotti}, {Harding}, {Labanti}, {Lapshov},
  {Lazzarotto}, {Lipari}, {Mauri}, {Morselli}, {Pacciani}, {Perotti},
  {Picozza}, {Prest}, {Pucella}, {Rapisarda}, {Rappoldi}, {Soffitta},
  {Trifoglio}, {Vallazza}, {Vercellone}, {Vittorini}, {Zambra}, {Zanello},
  {Pittori}, {Verrecchia}, {Preger}, {Santolamazza}, {Giommi}, \&
  {Salotti}}]{pel09a}
{Pellizzoni}, A., {Pilia}, M., {Possenti}, A., {et~al.} 2009{\natexlab{b}},
  \apj, 691, 1618

\bibitem[{{Pires}(2009)}]{pir09c}
{Pires}, A.~M. 2009, PhD thesis, ``Population study of radio-quiet and
  thermally emitting isolated neutron stars''; University of Strasbourg and
  University of Sao Paulo; 280 pages

\bibitem[{{Pires} {et~al.}(2009{\natexlab{a}}){Pires}, {Motch}, \&
  {Janot-Pacheco}}]{pir09b}
{Pires}, A.~M., {Motch}, C., \& {Janot-Pacheco}, E. 2009{\natexlab{a}}, \aap,
  504, 185

\bibitem[{{Pires} {et~al.}(2009{\natexlab{b}}){Pires}, {Motch}, {Turolla},
  {Treves}, \& {Popov}}]{pir09a}
{Pires}, A.~M., {Motch}, C., {Turolla}, R., {Treves}, A., \& {Popov}, S.~B.
  2009{\natexlab{b}}, \aap, 498, 233

\bibitem[{{Pittori} {et~al.}(2009){Pittori}, {Verrecchia}, {Chen},
  {Bulgarelli}, {Pellizzoni}, {Giuliani}, {Vercellone}, {Longo}, {Tavani},
  {Giommi}, {Barbiellini}, {Trifoglio}, {Gianotti}, {Argan}, {Antonelli},
  {Boffelli}, {Caraveo}, {Cattaneo}, {Cocco}, {Colafrancesco}, {Contessi},
  {Costa}, {Cutini}, {D'Ammando}, {Del Monte}, {de Paris}, {Di Cocco}, {di
  Persio}, {Donnarumma}, {Evangelista}, {Fanari}, {Feroci}, {Ferrari},
  {Fiorini}, {Fornari}, {Fuschino}, {Froysland}, {Frutti}, {Galli},
  {Gasparrini}, {Labanti}, {Lapshov}, {Lazzarotto}, {Liello}, {Lipari},
  {Mattaini}, {Marisaldi}, {Mastropietro}, {Mauri}, {Mauri}, {Mereghetti},
  {Morelli}, {Moretti}, {Morselli}, {Pacciani}, {Perotti}, {Piano}, {Picozza},
  {Pilia}, {Pontoni}, {Porrovecchio}, {Preger}, {Prest}, {Primavera},
  {Pucella}, {Rapisarda}, {Rappoldi}, {Rossi}, {Rubini}, {Sabatini},
  {Santolamazza}, {Scalise}, {Soffitta}, {Stellato}, {Striani}, {Tamburelli},
  {Traci}, {Trois}, {Vallazza}, {Vittorini}, {Zambra}, {Zanello}, \&
  {Salotti}}]{pit09}
{Pittori}, C., {Verrecchia}, F., {Chen}, A.~W., {et~al.} 2009, \aap, 506, 1563

\bibitem[{{Pons} {et~al.}(2009){Pons}, {Miralles}, \& {Geppert}}]{pon09}
{Pons}, J.~A., {Miralles}, J.~A., \& {Geppert}, U. 2009, \aap, 496, 207

\bibitem[{{Popov} {et~al.}(2003){Popov}, {Colpi}, {Prokhorov}, {Treves}, \&
  {Turolla}}]{pop03}
{Popov}, S.~B., {Colpi}, M., {Prokhorov}, M.~E., {Treves}, A., \& {Turolla}, R.
  2003, \aap, 406, 111

\bibitem[{{Popov} {et~al.}(2010){Popov}, {Pons}, {Miralles}, {Boldin}, \&
  {Posselt}}]{pop10}
{Popov}, S.~B., {Pons}, J.~A., {Miralles}, J.~A., {Boldin}, P.~A., \&
  {Posselt}, B. 2010, \mnras, 401, 2675

\bibitem[{{Posselt} {et~al.}(2007){Posselt}, {Popov}, {Haberl}, {Tr{\"u}mper},
  {Turolla}, \& {Neuh{\"a}user}}]{pos07}
{Posselt}, B., {Popov}, S.~B., {Haberl}, F., {et~al.} 2007, \apss, 308, 171

\bibitem[{{Posselt} {et~al.}(2008){Posselt}, {Popov}, {Haberl}, {Tr{\"u}mper},
  {Turolla}, \& {Neuh{\"a}user}}]{pos08}
{Posselt}, B., {Popov}, S.~B., {Haberl}, F., {et~al.} 2008, \aap, 482, 617

\bibitem[{{Predehl} {et~al.}(2010){Predehl}, {Andritschke}, {B{\"o}hringer},
  {Bornemann}, {Br{\"a}uninger}, {Brunner}, {Brusa}, {Burkert}, {Burwitz},
  {Cappelluti}, {Churazov}, {Dennerl}, {Eder}, {Elbs}, {Freyberg}, {Friedrich},
  {F{\"u}rmetz}, {Gaida}, {H{\"a}lker}, {Hartner}, {Hasinger}, {Hermann},
  {Huber}, {Kendziorra}, {von Kienlin}, {Kink}, {Kreykenbohm}, {Lamer},
  {Lapchov}, {Lehmann}, {Meidinger}, {Mican}, {Mohr}, {M{\"u}hlegger},
  {M{\"u}ller}, {Nandra}, {Pavlinsky}, {Pfeffermann}, {Reiprich}, {Robrade},
  {Roh{\'e}}, {Santangelo}, {Sch{\"a}chner}, {Schanz}, {Schmid}, {Schmitt},
  {Schreib}, {Schrey}, {Schwope}, {Steinmetz}, {Str{\"u}der}, {Sunyaev},
  {Tenzer}, {Tiedemann}, {Vongehr}, \& {Wilms}}]{pre10}
{Predehl}, P., {Andritschke}, R., {B{\"o}hringer}, H., {et~al.} 2010, in
  Society of Photo-Optical Instrumentation Engineers (SPIE) Conference Series,
  Vol. 7732, Society of Photo-Optical Instrumentation Engineers (SPIE)
  Conference Series

\bibitem[{{Predehl} \& {Schmitt}(1995)}]{pre95}
{Predehl}, P. \& {Schmitt}, J.~H.~M.~M. 1995, \aap, 293, 889

\bibitem[{{Raymond} \& {Smith}(1977)}]{ray77}
{Raymond}, J.~C. \& {Smith}, B.~W. 1977, \apjs, 35, 419

\bibitem[{{Reynolds} {et~al.}(2006){Reynolds}, {Borkowski}, {Gaensler}, {Rea},
  {McLaughlin}, {Possenti}, {Israel}, {Burgay}, {Camilo}, {Chatterjee},
  {Kramer}, {Lyne}, \& {Stairs}}]{rey06}
{Reynolds}, S.~P., {Borkowski}, K.~J., {Gaensler}, B.~M., {et~al.} 2006, \apjl,
  639, L71

\bibitem[{{Rutledge} {et~al.}(2008){Rutledge}, {Fox}, \& {Shevchuk}}]{rut08}
{Rutledge}, R.~E., {Fox}, D.~B., \& {Shevchuk}, A.~H. 2008, \apj, 672, 1137

\bibitem[{{Rutledge} {et~al.}(2003){Rutledge}, {Fox}, {Bogosavljevic}, \&
  {Mahabal}}]{rut03}
{Rutledge}, R.~E., {Fox}, D.~W., {Bogosavljevic}, M., \& {Mahabal}, A. 2003,
  \apj, 598, 458

\bibitem[{{Sanwal} {et~al.}(2002){Sanwal}, {Pavlov}, {Zavlin}, \&
  {Teter}}]{san02}
{Sanwal}, D., {Pavlov}, G.~G., {Zavlin}, V.~E., \& {Teter}, M.~A. 2002, \apjl,
  574, L61

\bibitem[{{Seward} {et~al.}(1979){Seward}, {Forman}, {Giacconi}, {Griffiths},
  {Harnden}, {Jones}, \& {Pye}}]{sew79}
{Seward}, F.~D., {Forman}, W.~R., {Giacconi}, R., {et~al.} 1979, \apjl, 234,
  L55

\bibitem[{{Shabaltas} \& {Lai}(2011)}]{sha11}
{Shabaltas}, N. \& {Lai}, D. 2011, ArXiv e-prints, astro-ph/1110.3129

\bibitem[{{Shternin} {et~al.}(2011){Shternin}, {Yakovlev}, {Heinke}, {Ho}, \&
  {Patnaude}}]{sht11}
{Shternin}, P.~S., {Yakovlev}, D.~G., {Heinke}, C.~O., {Ho}, W.~C.~G., \&
  {Patnaude}, D.~J. 2011, \mnras, 412, L108

\bibitem[{{Smith}(2006{\natexlab{a}})}]{smi06b}
{Smith}, N. 2006{\natexlab{a}}, \mnras, 367, 763

\bibitem[{{Smith}(2006{\natexlab{b}})}]{smi06a}
{Smith}, N. 2006{\natexlab{b}}, \apj, 644, 1151

\bibitem[{{Smith} {et~al.}(2010{\natexlab{a}}){Smith}, {Bally}, \&
  {Walborn}}]{smi10a}
{Smith}, N., {Bally}, J., \& {Walborn}, N.~R. 2010{\natexlab{a}}, \mnras, 405,
  1153

\bibitem[{{Smith} \& {Brooks}(2008)}]{smi08}
{Smith}, N. \& {Brooks}, K.~J. 2008, {The Carina Nebula: A Laboratory for
  Feedback and Triggered Star Formation}, ed. {Reipurth, B.}, 138

\bibitem[{{Smith} {et~al.}(2010{\natexlab{b}}){Smith}, {Povich}, {Whitney},
  {Churchwell}, {Babler}, {Meade}, {Bally}, {Gehrz}, {Robitaille}, \&
  {Stassun}}]{smi10b}
{Smith}, N., {Povich}, M.~S., {Whitney}, B.~A., {et~al.} 2010{\natexlab{b}},
  \mnras, 406, 952

\bibitem[{{Speagle} {et~al.}(2011){Speagle}, {Kaplan}, \& {van
  Kerkwijk}}]{spe11}
{Speagle}, J.~S., {Kaplan}, D.~L., \& {van Kerkwijk}, M.~H. 2011, \apj, 743,
  183

\bibitem[{{Stetson}(1987)}]{ste87}
{Stetson}, P.~B. 1987, \pasp, 99, 191

\bibitem[{{Str{\"u}der} {et~al.}(2001){Str{\"u}der}, {Briel}, {Dennerl},
  {Hartmann}, {Kendziorra}, {Meidinger}, {Pfeffermann}, {Reppin}, {Aschenbach},
  {Bornemann}, {Br{\"a}uninger}, {Burkert}, {Elender}, {Freyberg}, {Haberl},
  {Hartner}, {Heuschmann}, {Hippmann}, {Kastelic}, {Kemmer}, {Kettenring},
  {Kink}, {Krause}, {M{\"u}ller}, {Oppitz}, {Pietsch}, {Popp}, {Predehl},
  {Read}, {Stephan}, {St{\"o}tter}, {Tr{\"u}mper}, {Holl}, {Kemmer}, {Soltau},
  {St{\"o}tter}, {Weber}, {Weichert}, {von Zanthier}, {Carathanassis}, {Lutz},
  {Richter}, {Solc}, {B{\"o}ttcher}, {Kuster}, {Staubert}, {Abbey}, {Holland},
  {Turner}, {Balasini}, {Bignami}, {La Palombara}, {Villa}, {Buttler},
  {Gianini}, {Lain{\'e}}, {Lumb}, \& {Dhez}}]{str01}
{Str{\"u}der}, L., {Briel}, U., {Dennerl}, K., {et~al.} 2001, \aap, 365, L18

\bibitem[{{Tavani} {et~al.}(2009){Tavani}, {Barbiellini}, {Argan}, {Boffelli},
  {Bulgarelli}, {Caraveo}, {Cattaneo}, {Chen}, {Cocco}, {Costa}, {D'Ammando},
  {Del Monte}, {de Paris}, {Di Cocco}, {di Persio}, {Donnarumma},
  {Evangelista}, {Feroci}, {Ferrari}, {Fiorini}, {Fornari}, {Fuschino},
  {Froysland}, {Frutti}, {Galli}, {Gianotti}, {Giuliani}, {Labanti}, {Lapshov},
  {Lazzarotto}, {Liello}, {Lipari}, {Longo}, {Mattaini}, {Marisaldi},
  {Mastropietro}, {Mauri}, {Mauri}, {Mereghetti}, {Morelli}, {Morselli},
  {Pacciani}, {Pellizzoni}, {Perotti}, {Piano}, {Picozza}, {Pontoni},
  {Porrovecchio}, {Prest}, {Pucella}, {Rapisarda}, {Rappoldi}, {Rossi},
  {Rubini}, {Soffitta}, {Traci}, {Trifoglio}, {Trois}, {Vallazza},
  {Vercellone}, {Vittorini}, {Zambra}, {Zanello}, {Pittori}, {Preger},
  {Santolamazza}, {Verrecchia}, {Giommi}, {Colafrancesco}, {Antonelli},
  {Cutini}, {Gasparrini}, {Stellato}, {Fanari}, {Primavera}, {Tamburelli},
  {Viola}, {Guarrera}, {Salotti}, {D'Amico}, {Marchetti}, {Crisconio},
  {Sabatini}, {Annoni}, {Alia}, {Longoni}, {Sanquerin}, {Battilana}, {Concari},
  {Dessimone}, {Grossi}, {Parise}, {Monzani}, {Artina}, {Pavesi},
  {Marseguerra}, {Nicolini}, {Scandelli}, {Soli}, {Vettorello}, {Zardetto},
  {Bonati}, {Maltecca}, {D'Alba}, {Patan{\'e}}, {Babini}, {Onorati},
  {Acquaroli}, {Angelucci}, {Morelli}, {Agostara}, {Cerone}, {Michetti},
  {Tempesta}, {D'Eramo}, {Rocca}, {Giannini}, {Borghi}, {Garavelli}, {Conte},
  {Balasini}, {Ferrario}, {Vanotti}, {Collavo}, \& {Giacomazzo}}]{tav09}
{Tavani}, M., {Barbiellini}, G., {Argan}, A., {et~al.} 2009, \aap, 502, 995

\bibitem[{{The Fermi-LAT Collaboration}(2011)}]{2fgl11}
{The Fermi-LAT Collaboration}. 2011, ArXiv e-prints, astro-ph/1108.1435

\bibitem[{{Tody}(1986)}]{tod86}
{Tody}, D. 1986, in Society of Photo-Optical Instrumentation Engineers (SPIE)
  Conference Series, Vol. 627, Society of Photo-Optical Instrumentation
  Engineers (SPIE) Conference Series, ed. {D.~L.~Crawford}, 733--+

\bibitem[{{Townsley} {et~al.}(2011{\natexlab{a}}){Townsley}, {Broos}, {Chu},
  {Gagn{\'e}}, {Garmire}, {Gruendl}, {Hamaguchi}, {Mac Low}, {Montmerle},
  {Naz{\'e}}, {Oey}, {Park}, {Petre}, \& {Pittard}}]{tow11b}
{Townsley}, L.~K., {Broos}, P.~S., {Chu}, Y.-H., {et~al.} 2011{\natexlab{a}},
  \apjs, 194, 15

\bibitem[{{Townsley} {et~al.}(2011{\natexlab{b}}){Townsley}, {Broos},
  {Corcoran}, {Feigelson}, {Gagn{\'e}}, {Montmerle}, {Oey}, {Smith}, {Garmire},
  {Getman}, {Povich}, {Remage Evans}, {Naz{\'e}}, {Parkin}, {Preibisch},
  {Wang}, {Wolk}, {Chu}, {Cohen}, {Gruendl}, {Hamaguchi}, {King}, {Mac Low},
  {McCaughrean}, {Moffat}, {Oskinova}, {Pittard}, {Stassun}, {ud-Doula},
  {Walborn}, {Waldron}, {Churchwell}, {Nichols}, {Owocki}, \&
  {Schulz}}]{tow11a}
{Townsley}, L.~K., {Broos}, P.~S., {Corcoran}, M.~F., {et~al.}
  2011{\natexlab{b}}, \apjs, 194, 1

\bibitem[{{Turner} {et~al.}(2001){Turner}, {Abbey}, {Arnaud}, {Balasini},
  {Barbera}, {Belsole}, {Bennie}, {Bernard}, {Bignami}, {Boer}, {Briel},
  {Butler}, {Cara}, {Chabaud}, {Cole}, {Collura}, {Conte}, {Cros}, {Denby},
  {Dhez}, {Di Coco}, {Dowson}, {Ferrando}, {Ghizzardi}, {Gianotti}, {Goodall},
  {Gretton}, {Griffiths}, {Hainaut}, {Hochedez}, {Holland}, {Jourdain},
  {Kendziorra}, {Lagostina}, {Laine}, {La Palombara}, {Lortholary}, {Lumb},
  {Marty}, {Molendi}, {Pigot}, {Poindron}, {Pounds}, {Reeves}, {Reppin},
  {Rothenflug}, {Salvetat}, {Sauvageot}, {Schmitt}, {Sembay}, {Short},
  {Spragg}, {Stephen}, {Str{\"u}der}, {Tiengo}, {Trifoglio}, {Tr{\"u}mper},
  {Vercellone}, {Vigroux}, {Villa}, {Ward}, {Whitehead}, \& {Zonca}}]{tur01}
{Turner}, M.~J.~L., {Abbey}, A., {Arnaud}, M., {et~al.} 2001, \aap, 365, L27

\bibitem[{{Turner} {et~al.}(2010){Turner}, {Rutledge}, {Letcavage}, {Shevchuk},
  \& {Fox}}]{tur10}
{Turner}, M.~L., {Rutledge}, R.~E., {Letcavage}, R., {Shevchuk}, A.~S.~H., \&
  {Fox}, D.~B. 2010, \apj, 714, 1424

\bibitem[{{Turolla}(2009)}]{tur09}
{Turolla}, R. 2009, in Astrophysics and Space Science Library, Vol. 357,
  Astrophysics and Space Science Library, ed. W.~{Becker}, 141--163

\bibitem[{{Turolla} {et~al.}(2011){Turolla}, {Zane}, {Pons}, {Esposito}, \&
  {Rea}}]{tur11}
{Turolla}, R., {Zane}, S., {Pons}, J.~A., {Esposito}, P., \& {Rea}, N. 2011,
  \apj, 740, 105

\bibitem[{{van Kerkwijk} \& {Kaplan}(2007)}]{ker07a}
{van Kerkwijk}, M.~H. \& {Kaplan}, D.~L. 2007, \apss, 308, 191

\bibitem[{{van Kerkwijk} \& {Kulkarni}(2001)}]{ker01}
{van Kerkwijk}, M.~H. \& {Kulkarni}, S.~R. 2001, \aap, 380, 221

\bibitem[{{Voges} {et~al.}(1999){Voges}, {Aschenbach}, {Boller},
  {Br{\"a}uninger}, {Briel}, {Burkert}, {Dennerl}, {Englhauser}, {Gruber},
  {Haberl}, {Hartner}, {Hasinger}, {K{\"u}rster}, {Pfeffermann}, {Pietsch},
  {Predehl}, {Rosso}, {Schmitt}, {Tr{\"u}mper}, \& {Zimmermann}}]{vog99b}
{Voges}, W., {Aschenbach}, B., {Boller}, T., {et~al.} 1999, \aap, 349, 389

\bibitem[{{Walborn} {et~al.}(2007){Walborn}, {Smith}, {Howarth}, {Vieira
  Kober}, {Gull}, \& {Morse}}]{wal07}
{Walborn}, N.~R., {Smith}, N., {Howarth}, I.~D., {et~al.} 2007, \pasp, 119, 156

\bibitem[{{Wang} {et~al.}(2011){Wang}, {Feigelson}, {Townsley}, {Broos},
  {Getman}, {Wolk}, {Preibisch}, {Stassun}, {Moffat}, {Garmire}, {King},
  {McCaughrean}, \& {Zinnecker}}]{wan11}
{Wang}, J., {Feigelson}, E.~D., {Townsley}, L.~K., {et~al.} 2011, \apjs, 194,
  11

\bibitem[{{Wilms} {et~al.}(2000){Wilms}, {Allen}, \& {McCray}}]{wil00}
{Wilms}, J., {Allen}, A., \& {McCray}, R. 2000, \apj, 542, 914

\bibitem[{{Yakovlev} \& {Pethick}(2004)}]{yak04}
{Yakovlev}, D.~G. \& {Pethick}, C.~J. 2004, \araa, 42, 169

\bibitem[{{Zane} {et~al.}(2011){Zane}, {Haberl}, {Israel}, {Pellizzoni},
  {Burgay}, {Mignani}, {Turolla}, {Possenti}, {Esposito}, {Champion},
  {Eatough}, {Barr}, \& {Kramer}}]{zan11}
{Zane}, S., {Haberl}, F., {Israel}, G.~L., {et~al.} 2011, \mnras, 410, 2428

\bibitem[{{Zavlin} {et~al.}(1996){Zavlin}, {Pavlov}, \& {Shibanov}}]{zav96}
{Zavlin}, V.~E., {Pavlov}, G.~G., \& {Shibanov}, Y.~A. 1996, \aap, 315, 141

\end{thebibliography}
\end{document}